\PassOptionsToPackage{section,below}{placeins}
\documentclass[a4paper]{bytedance_seed}

\usepackage[utf8]{inputenc}
\usepackage{url}
\usepackage{booktabs}
\usepackage{amsfonts}
\usepackage{amsmath}
\usepackage{amssymb}
\usepackage{nicefrac}
\usepackage{graphicx}
\usepackage{xcolor}
\usepackage{multirow}
\usepackage{float}
\usepackage{tabularx}
\usepackage{oridynamic_style}

\ifpdftex
  \usepackage{tgpagella}
  \usepackage[T1]{fontenc}
\else
  \usepackage{unicode-math}
\fi

\hypersetup{citecolor=black}
\graphicspath{{figs/}}

\input{numbers}

\DeclareTextFontCommand{\textbf}{\bfseries}
\newcommand{\calpha}{Calpha}

\title{\raisebox{-0.22em}{\includegraphics[height=1.25em]{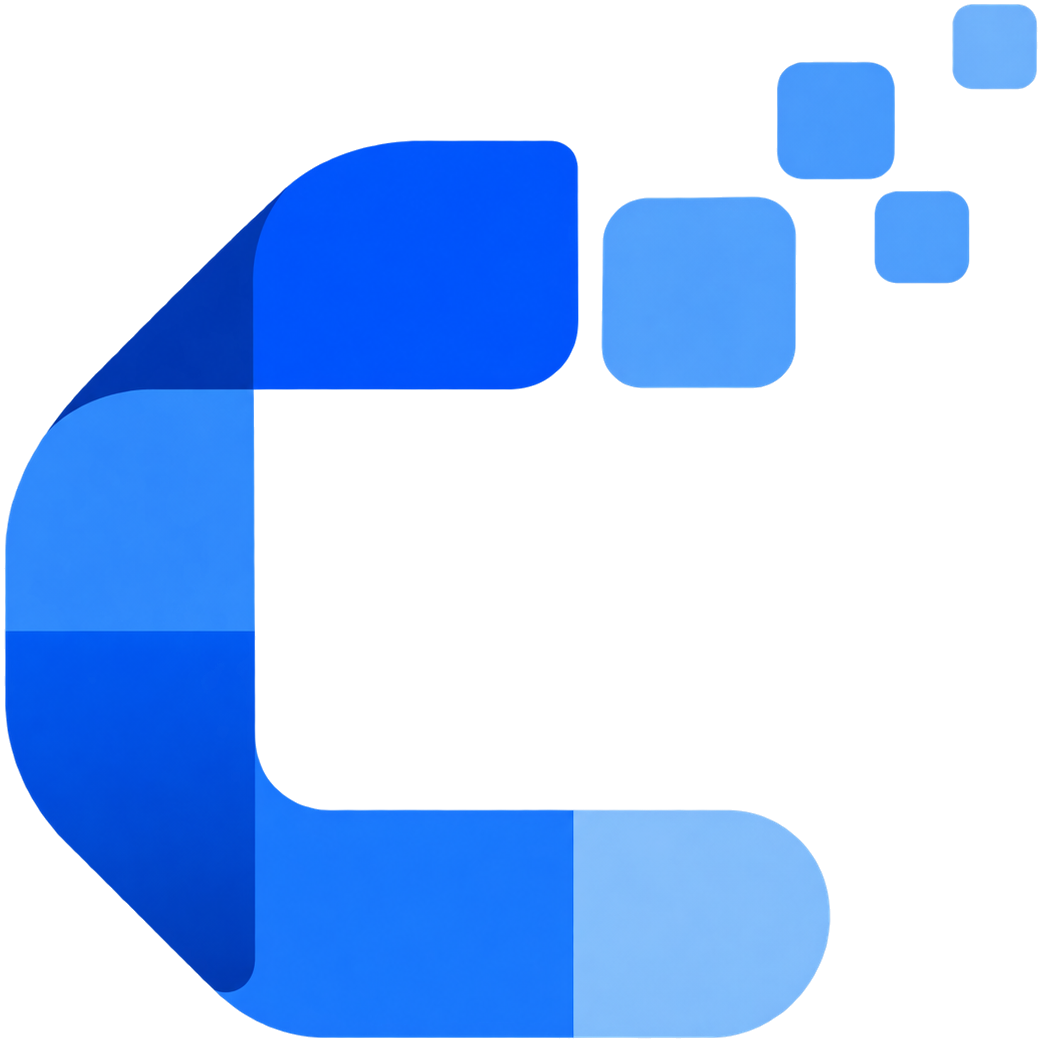}}\hspace{0.3em}\calpha{}: Toward Native Biomolecular Intelligence\\
       in a General-Purpose Language Model\\[5pt]
       {\fontsize{12}{14}\selectfont Technical Report I: Protein Folding}}

\author[1,2]{Yong Liu}
\author[3,4]{Zhanpeng Shi}
\author[2]{Peicong Lin}
\author[1]{Zhenyi Zhong}
\author[5]{Yizhou Dang}
\author[1]{Wendi Wu}
\author[6]{Haoling Li}
\author[1,2]{Shuangjia Zheng}
\affiliation[1]{Shanghai Jiao Tong University, Shanghai, China}
\affiliation[2]{Lingang Laboratory, Shanghai, China}
\affiliation[3]{Fudan University, Shanghai, China}
\affiliation[4]{Shanghai Innovation Institute, Shanghai, China}
\affiliation[5]{Northeastern University, Shenyang, China}
\affiliation[6]{Nanjing University, Nanjing, China}

\newcommand{\calphaemail}{liuyong9975@sjtu.edu.cn}
\newcommand{\calphahomepage}{https://calpha-ai.com}
\newcommand{\calphagithub}{https://github.com/yliust/Calpha}
\newcommand{\calphahf}{https://huggingface.co/liuyong9975/Calpha}
\newcommand{\linkicon}[1]{\raisebox{-0.14em}{\includegraphics[height=0.95em]{#1}}}
\newcommand{\linkrow}[3]{
  \makebox[1.2em][c]{#1}\enspace\textbf{\sffamily #2} &
  \ifdefempty{#3}{}{\href{#3}{\textcolor{black!65}{\nolinkurl{#3}}}}}
\renewcommand{\checkdatalist}{%
  \begingroup
    \small
    \renewcommand{\UrlFont}{\fontfamily{bytesansmedium}\selectfont}%
    \vspace{1mm}
    \noindent
    \begin{tabularx}{\linewidth}{@{}l@{\hspace{1.8em}}>{\footnotesize\raggedright\arraybackslash}X@{}}
    \makebox[1.2em][c]{\linkicon{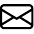}}\enspace\textbf{\sffamily Correspondence} &
      \href{mailto:\calphaemail}{\textcolor{black!65}{\nolinkurl{\calphaemail}}} \\[4pt]
    \linkrow{\linkicon{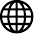}}{Homepage}{\calphahomepage} \\[4pt]
    \linkrow{\linkicon{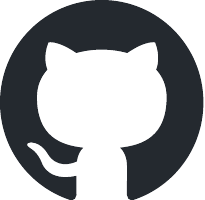}}{GitHub}{\calphagithub} \\[4pt]
    \linkrow{\raisebox{-0.18em}{\includegraphics[height=1.05em]{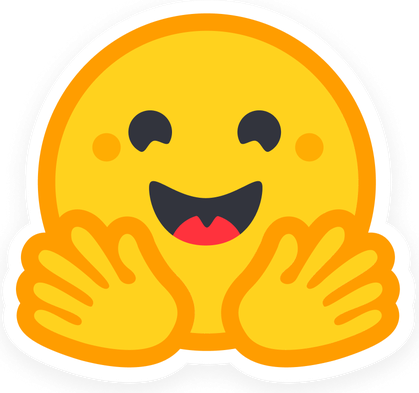}}}{Hugging Face}{\calphahf}
    \end{tabularx}\par
  \endgroup
}

\abstract{%
Biomolecular foundation models predict the structures of proteins and their complexes
accurately, and large language models (LLMs) reason in natural language, but the two
remain separate: the abilities of the former are not part of a general-purpose LLM, and
the reasoning of the latter does not enter a structure prediction.
\calpha{} aims at one general model that folds proteins and complexes, understands
protein sequence and structure, and lets its language reasoning influence the structures
it predicts. This report covers the first step, in which an LLM outputs single-chain
protein structures.
Qwen3.5-9B reads a sequence enriched with ESM2 residue features. It can respond in text
or emit a token that invokes a pretrained SimpleFold structure head, which receives
information about the protein only through Qwen's residue states.
Under a fixed folding request, the head achieves mean TM-scores of
\N{ours_b2.foldbench334.overall} on the FoldBench-334 development set,
\N{ours_b2.casp16.overall} on CASP16 domains and \N{ours_b2.cameo2025.overall} on a
CAMEO-style PDB set. We compare these predictions with a baseline in which ESM2-3B
encodes the sequence and a fine-tuned SimpleFold-3B head predicts the coordinates. The
two systems also differ in trainable weights, training duration and crop length. For
chains of 100--299 residues, all three paired confidence intervals include zero. At
300--600 residues, the mean differences are
\N{ours_b2__vs__esm2_direct.foldbench334.len300-600.delta},
\N{ours_b2__vs__esm2_direct.casp16.len300-600.delta} and
\N{ours_b2__vs__esm2_direct.cameo2025.len300-600.delta}, respectively.
In a separate test, four English request types triggered folding on every evaluated
example. Protein training reduced accuracy on a 350-item MMLU-Pro subset from 77.4\% to
70.3\%.
An LLM can therefore output single-chain structures from its own states, with lower long-chain
scores in the current comparisons and reduced accuracy on the evaluated MMLU-Pro subset. Complex prediction, structure
understanding and the effect of language reasoning on the predicted coordinates remain
future work.
}

\AtBeginDocument{\renewcommand{\bffont}{}\setstretch{1.05}\setlength{\parskip}{4pt plus 1pt minus 1pt}}
\titlespacing*{\section}{0pt}{15pt plus 3pt minus 2pt}{7pt}
\titlespacing*{\subsection}{0pt}{11pt plus 2pt minus 1pt}{5pt}
\titlespacing*{\subsubsection}{0pt}{9pt plus 2pt minus 1pt}{4pt}
\begin{document}

\maketitle
\clearpage

\section{Introduction}

\paragraph{Foundation models for biomolecular structure.}
Predicting the three-dimensional structure of a protein from its sequence~\cite{anfinsen1973}
is now a routine computation. AlphaFold~\cite{senior2020alphafold,jumper2021af2} and
RoseTTAFold~\cite{baek2021rosettafold} predict single chains from multiple sequence
alignments, in many cases with accuracy comparable to experiment.
ESMFold~\cite{lin2023esmfold} and OmegaFold~\cite{wu2022omegafold} replace the alignment
with the representations of a protein language model trained on sequences
alone~\cite{rives2021esm1b,elnaggar2022prottrans}. Prediction has since moved from single
chains to assemblies: AlphaFold-Multimer~\cite{evans2021afmultimer} predicts protein
complexes, and AlphaFold~3~\cite{abramson2024af3}, RoseTTAFold
All-Atom~\cite{krishna2024rfaa}, Chai-1~\cite{chai2024chai1},
Boltz~\cite{wohlwend2024boltz1,passaro2025boltz2} and OpenDDE~\cite{opendde} model proteins together with nucleic
acids and small molecules. The architectures have also become simpler and generative.
AlphaFlow~\cite{jing2024alphaflow} turns a structure predictor into a flow-matching
model~\cite{lipman2023flow}, and SimpleFold~\cite{wang2025simplefold} folds proteins with a
standard transformer trained by flow matching. Related models design
backbones~\cite{watson2023rfdiffusion,ingraham2023chroma}, design sequences for a given
backbone~\cite{dauparas2022proteinmpnn}, and model sequence, structure and function
jointly~\cite{hayes2025esm3,wang2025dplm2}. These models now serve as foundation models for
structural biology. Each of them, however, maps a fixed kind of input to a fixed kind of
output. None takes an instruction in natural language, uses what is written about a
protein, or explains its answer.

\paragraph{Large language models.}
Large language models (LLMs) have developed along a different path. Transformers trained on
text at scale~\cite{vaswani2017attention,brown2020gpt3,openai2023gpt4,yang2025qwen3} follow
instructions, answer questions across scientific disciplines, and reason step by step in
natural language~\cite{wei2022cot,guo2025deepseekr1}. Equipped with
tools~\cite{yao2023react}, they plan and carry out multi-step research tasks in
chemistry~\cite{boiko2023coscientist,bran2024chemcrow} and
biology~\cite{gottweis2025coscientist,swanson2025virtuallab,huang2025biomni}. They do not,
however, produce molecular structure by themselves. When we ask four frontier chat models
to write the C$\alpha$ coordinates of a protein without tools, their mean TM-scores lie
between \N{llm_deepseek.cameo2025.subset} and \N{llm_gpt.casp16.subset}, against
\N{floor_randomwalk.cameo2025.subset}--\N{floor_randomwalk.foldbench334.subset} for a
random-walk reference on the same targets (Section~\ref{sec:q1}).

\begin{figure}[t]
\centering
\includegraphics[width=\linewidth]{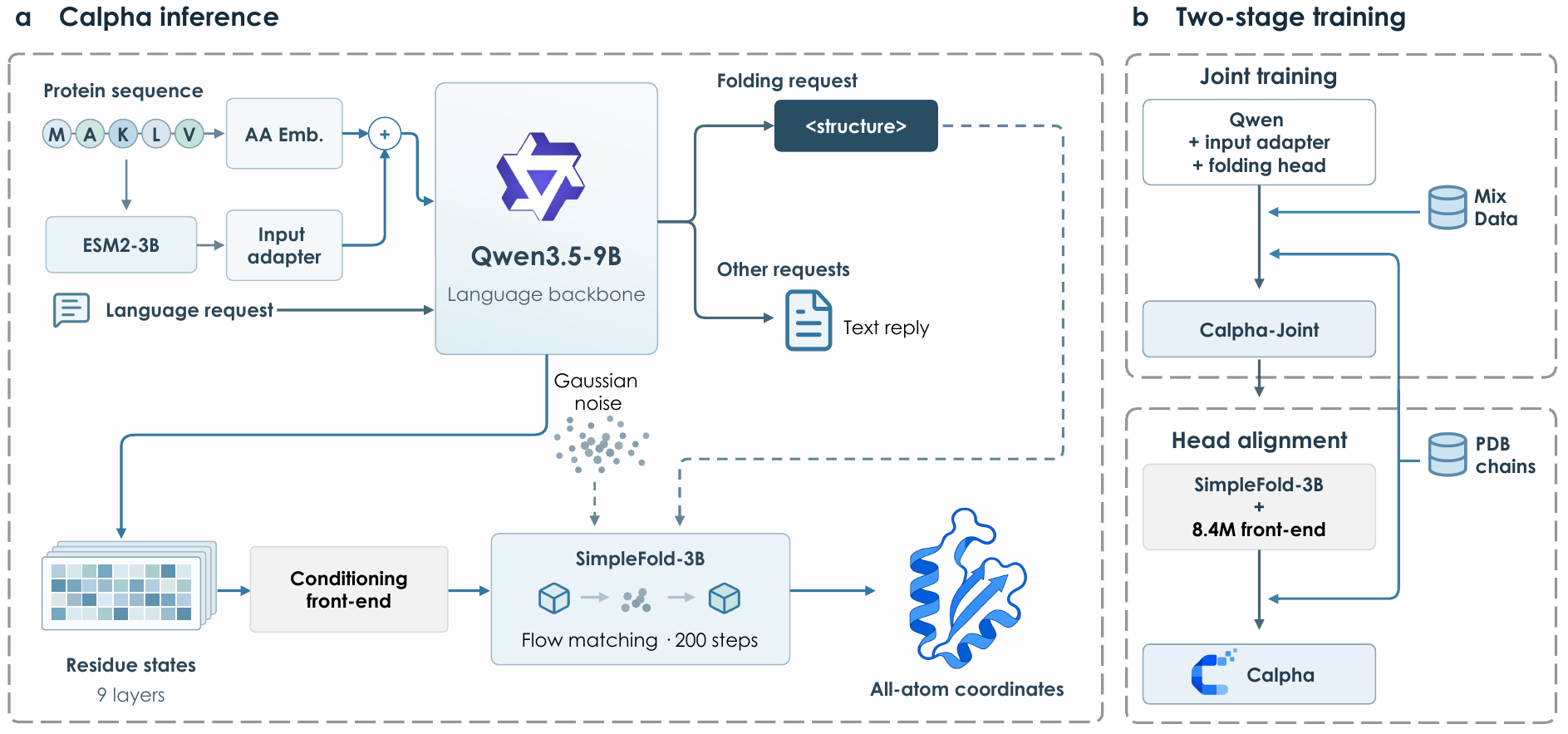}
\caption{\calpha{} inference and training. \textbf{a} Frozen ESM2 features enter Qwen at
the amino-acid positions. When Qwen emits \texttt{<structure>}, the front-end mixes residue
states from nine layers to condition SimpleFold-3B. \textbf{b} Joint training updates Qwen,
the input adapter and a 100M folding head. The second stage freezes Qwen and the pretrained
3B head and trains the 8.4M-parameter conditioning front-end.}
\label{fig:arch}
\end{figure}

\paragraph{The gap between the two.}
The abilities of biomolecular foundation models are therefore important to science but are
not yet part of general-purpose LLMs. Existing work connects the two in three ways. In the
first, an LLM agent calls a folding or design program as an external tool and reads back
its
output~\cite{ghafarollahi2024protagents,huang2025biomni,swanson2025virtuallab,ge2026autobinder}.
The language model chooses what to run, but its reasoning stays outside the predictor and
can affect a prediction only through the predictor's fixed inputs. In the second, a protein
encoder is attached to an LLM so that the model can answer questions about a sequence or a
structure in
text~\cite{xu2023protst,zhuo2024protllm,xiao2024proteingpt,wang2024protchatgpt,zhou2025evolla},
following vision--language models~\cite{alayrac2022flamingo,li2023blip2,liu2023llava}.
These models describe proteins and do not generate coordinates. In the third, structure is
converted into discrete tokens that a language model reads and
writes~\cite{chen2024helixprotx,ma2025prottex,pei2026biomatrix,lu2026trion}, or a
continuous generator is attached to a sequence
model~\cite{zhang2025unigenx,zhou2025hdprot}, as generative heads are attached to language
models for images and robot
actions~\cite{li2024mar,zhou2024transfusion,black2024pi0}. In all three, the reasoning that
an LLM carries out in language has not been shown to improve the structure that is
predicted.

\paragraph{Our goal.}
\calpha{} is a project to build a general-purpose LLM in which these abilities are native.
One model should fold proteins and predict complexes, and it should understand protein
sequence and structure well enough to discuss them. The abilities that today belong to
separate biomolecular foundation models would then be part of a general model and connected
to everything else that it knows. Most importantly, we want the model's natural-language
reasoning to take part in the prediction. What the model reads in the literature, what a
scientist tells it and what it infers about a mechanism should influence, and in some cases
determine, the structure it predicts. A model of this kind could be asked for the bound
state of a protein, or could revise a prediction after reasoning about why it is wrong. We
see this as a route to better scientific discovery with LLMs. This goal has not been
reached. Section~\ref{sec:road} lays out the roadmap, and this report covers its first
step.

\paragraph{This report: single-chain structures from an LLM.}
The first step is to make an LLM output the structure of a single protein chain. In
\calpha{} (Figure~\ref{fig:arch}), Qwen3.5-9B~\cite{qwen35} reads a protein sequence
inside an ordinary chat turn, together with residue features from
ESM2-3B~\cite{lin2023esmfold}. For a question about the protein, it replies in text. For a
folding request, it emits a \texttt{<structure>} token, and its hidden states at the
residue positions condition a pretrained SimpleFold-3B flow-matching
head~\cite{wang2025simplefold} that generates all-atom coordinates. The head receives
information about the protein only through the language model. The same forward pass
therefore decides whether to fold and supplies what the head folds from, with no tool
call. The question we can answer at this stage is how much folding accuracy is retained
when the structure head receives its input through the chat model.

\paragraph{What we find.}
We report three observations. First, the states exported from Qwen support single-chain
folding: the head reaches mean TM-scores of \N{ours_b2.foldbench334.overall} on our
development set, \N{ours_b2.casp16.overall} on CASP16 domains and
\N{ours_b2.cameo2025.overall} on a CAMEO-style set. Second, the current \calpha{} configuration scores lower than the direct
ESM2 $\rightarrow$ SimpleFold-3B baseline on chains of 300--600 residues in all three sets.
At 100--299 residues, all three paired confidence intervals include zero. The two systems also differ in which weights they update, in training duration and
in crop length (Table~\ref{tab:recipes}), so the comparison measures their combined
effect. Third, four of the five tested folding request types elicited the trigger on every
evaluated sequence. Accuracy on the evaluated MMLU-Pro subset~\cite{wang2024mmlupro}
fell by 7.1 percentage points after protein training.

\paragraph{What remains open.}
These results concern single chains scored on C$\alpha$ atoms under one fixed request.
\calpha{} does not yet predict complexes, and it does not yet take structure as input.
In none of our comparisons does routing through the language model make folding more
accurate than the direct ESM2 baseline. Whether a change
of wording preserves coordinate accuracy, and whether an instruction or a line of reasoning
can steer the predicted conformation, are the questions that the goal above depends on.
They are open (Section~\ref{sec:limits}).

\section{System and training}
\label{sec:system}

\calpha{} has four parts that run at inference: a protein encoder, a chat model, a small
conditioning front-end and a structure head.
Two of them, the encoder and the head, are pretrained specialist models used with their public
weights.
We first describe the inference path, then the two training stages, and Table~\ref{tab:stages}
states for every component where its weights come from and when it is trained.

\subsection{Inference path}

\paragraph{Input.}
The vocabulary of Qwen3.5-9B~\cite{qwen35} is extended with amino-acid tokens and a
\texttt{<structure>} token.
A protein is written as a span of amino-acid tokens inside an ordinary chat turn.
Frozen ESM2-3B~\cite{lin2023esmfold} computes per-residue features, which a small adapter (a
learned mixture over ESM2 layers, an MLP and a zero-initialised gate) maps to the model width
and adds to the embeddings of the corresponding amino-acid tokens.
The language model therefore sees each residue as ``its letter plus what ESM2 knows about it''.
ESM2 is thus a second model that reads the sequence; what the design guarantees is that its
features reach the structure head only through the language model.

\paragraph{Trigger.}
The language head is unchanged.
Emitting the token \texttt{<structure>} invokes the structure head; otherwise the
language model continues generating text. Section~\ref{sec:onemodel} measures this decision.
Whether to fold is thus decided by the model itself, in the same forward pass as any other
answer.

\paragraph{Conditioning front-end.}
After the trigger we read the hidden states at the residue positions from nine layers of the
32-layer model (layers $0, 4, \dots, 32$; width 4096).
This gives a tensor of shape $(L, 9, 4096)$, the same form as the per-layer ESM2 tensor that
SimpleFold consumes.
The front-end is SimpleFold's own conditioning entry, rebuilt for this shape: a learned softmax
mixture over the nine layers followed by a projection into the head.
It has 8.4M parameters.

\paragraph{Structure head.}
The head is SimpleFold-3B (2.86B parameters) with its public weights: an atom encoder, a residue
trunk and an atom decoder trained with flow matching to transport Gaussian noise to all-atom
coordinates.
We sample with 200 steps and the default noise scale ($\tau = 0.01$) and report a single sample
per target.
The head outputs all atoms, but every accuracy number in this report is computed on alpha
carbons only; side-chain accuracy was not evaluated.

\subsection{Training}

\paragraph{Stage 1: teach the language model to read proteins and to decide when to fold.}
All parameters of Qwen3.5-9B, the ESM2 adapter and a small SimpleFold-100M head, initialised
from its public weights, are trained jointly for 2B counted tokens.
The loss is the sum of three terms: the language cross-entropy, the cross-entropy of the
trigger token, and the head's flow-matching loss with a smooth-lDDT term, the last weighted by
28.9 to match gradient scales.
The language side mixes 75\% protein tasks (inverse folding, secondary structure, contact prediction, infilling) with 25\%
general text and reasoning data, all in English; the structure side uses 84{,}658 PDB chains
cropped to 256 residues.
The stage-1 model with its 100M head is itself a working folding model and we report it as
\calpha{}-Joint.
In trained parameters this stage is by far the larger of the two.

\paragraph{Stage 2: attach a large pretrained head and train only the front-end.}
The language model is frozen and exports hidden states.
The 100M head is replaced by SimpleFold-3B, which is also frozen; only the rebuilt front-end is
trained, for 20{,}000 steps at an effective batch size of 64 (learning rate $10^{-4}$, EMA
0.999) on the same 84{,}658 chains, followed by a further 10{,}000 steps with the crop length
raised from 256 to 512 residues.
The whole stage took about 11 hours on eight H200 GPUs.
The figure of 8.4M trained parameters describes this stage only: the language model whose
hidden states it reads was trained in full in stage~1.

\begin{table}[t]
\centering
\caption{Components of \calpha{}: where the weights come from, what is trained in each stage,
and what runs at inference. A dash means the component is not present in that stage.}
\label{tab:stages}
\small
\begin{tabular}{@{}llllc@{}}
\toprule
Component (parameters) & Initial weights & Stage 1 & Stage 2 & At inference \\
\midrule
ESM2-3B encoder & public & frozen & frozen & yes \\
Adapter, ESM2 to language model & new & trained & frozen & yes \\
Qwen3.5-9B ($\approx$9B) & public & trained & frozen & yes \\
SimpleFold-100M head (0.1B) & public & trained & -- & \calpha{}-Joint only \\
Conditioning front-end (8.4M) & new & -- & trained & yes \\
SimpleFold-3B head (2.86B) & public & -- & frozen & yes \\
\bottomrule
\end{tabular}
\end{table}

\subsection{The two systems compared on every target}

In our ESM2 $\rightarrow$ SimpleFold-3B baseline, ESM2-3B encodes the sequence into residue features.
A fine-tuned SimpleFold-3B head takes those features directly and generates the atomic coordinates.
In \calpha{}, the ESM2 features first pass through Qwen. A trained projection maps Qwen's residue states
to the inputs of a frozen SimpleFold-3B head. Both systems start from the same pretrained head weights
and use the same 84,658 training chains.
Table~\ref{tab:recipes} specifies the checkpoints used for evaluation.

\begin{table}[ht]
\centering\small
\caption{Training configurations for the final paired comparison. Crop is measured in residues.
Step counts are cumulative within the structure-head training run.}
\label{tab:recipes}
\begin{tabularx}{\linewidth}{@{}lXX@{}}
\toprule
 & \calpha{} & ESM2 $\rightarrow$ SimpleFold-3B \\
\midrule
Conditioning & Nine layers of Qwen residue states & ESM2 residue features \\
Updated weights & 8.4M conditioning front-end; frozen 3B head & Fine-tuned 3B head \\
Schedule & 20,000 steps at crop 256, then 10,000 at crop 512 & 10,000 steps at crop 256 \\
Evaluation checkpoint & EMA at step 30,000 & EMA at step 10,000 \\
\bottomrule
\end{tabularx}
\end{table}

Table~\ref{tab:main} also compares two variants. \calpha{}-Joint is the stage-1 model
with its co-trained 100M head. In \calpha{}-Token, Qwen writes DPLM-2 structure
tokens, which the corresponding decoder converts to coordinates~\cite{companion}. The public SimpleFold-3B baseline uses the
released weights.

\section{What we measured}
\label{sec:setup}

\begin{table}[ht]
\centering\small
\caption{Each measurement follows a different part of the system.}
\label{tab:measurement}
\begin{tabularx}{\linewidth}{@{}p{2.5cm}XX@{}}
\toprule
Test & Input and population & Recorded outcome \\
\midrule
Folding & Fixed request template; 334 development chains, 155 CASP16 domains and 200 PDB chains
& C$\alpha$ coordinates sampled offline from exported states; mean TM-score including failures \\
Request trigger & Sequences eligible under the length filter; six English request or question types
& Fraction of responses emitting \texttt{<structure>} for each type \\
General ability & Same 350 MMLU-Pro items before and after protein training & Five-shot accuracy \\
\bottomrule
\end{tabularx}
\end{table}

\paragraph{Three sets with different roles.}
FoldBench-334 contains chains released from 18 January 2023 to 23 October 2024. All development
choices were made on this set. CASP16 comprises 155 domain entries corresponding to 72 distinct
sequences. We predict domains in isolation, which differs from the official whole-target protocol.
Our third set contains 200 single-chain X-ray structures released by the PDB between 1 January
and 19 February 2025. We selected them directly from RCSB using CAMEO-style criteria when the
CAMEO service was unavailable~\cite{robin2021cameo}. Tables abbreviate this set as CAMEO-2025.
Nearest-training-structure retrieval already scores \N{retrieval.cameo2025.overall} on it.
We report the three sets separately throughout.

\paragraph{Homology buckets.}
Every target is searched against one training pool of 667{,}331 chains (PDB~\cite{berman2000pdb} plus AFDB-SwissProt~\cite{varadi2022afdb})
and classified using a minimum sequence identity of 30\% and coverage of at least 50\% on either side, then assigned to a \emph{homolog} or \emph{no-homolog} bucket.
The split is 206/128 for FoldBench-334, 85/70 for CASP16 and 184/16 for CAMEO-2025; the last
no-homolog bucket is too small to support conclusions on its own.
The buckets do not mean quite the same thing across sets: FoldBench-334 homologs were removed
from the data used to train our language model, whereas for the other two sets no such filtering
was done.
We also report four length buckets ($<$100, 100--299, 300--600, $>$600 residues).

\paragraph{Scoring and sampling.}
Our primary measure is TM-score~\cite{tmscore} with fixed residue correspondence on alpha
carbons. The scorer also records lDDT~\cite{lddt}, GDT-TS and RMSD. Failed predictions receive
zero TM. We report completion separately. For \calpha{} the primary result uses one sample
per target; Table~\ref{tab:multisample} examines 16. Calpha-Token uses per-target
scores computed earlier with the same metric implementation. OpenDDE~\cite{opendde} uses its default top-ranked
sample of five. Its MSA-assisted row uses an external alignment and is shown for context.
Random walks, nearest-training-structure retrieval and experimental coordinates give three
references. The reported geometry check measures adjacent C$\alpha$ spacing. Scorer outputs
also include nonlocal C$\alpha$ proximity and normalized radius of gyration where available;
none of these checks validates side-chain geometry.

\paragraph{Paired comparisons.}
For each target we subtract the reference system's TM-score from \calpha{}'s. We report the
mean difference and a percentile 95\% interval from 10,000 paired bootstrap resamples of
entries. The intervals quantify variation across entries for one trained model per
configuration. Related sequences and multiple CASP entries from a single target can reduce
the effective sample size. Interval estimates condition on these training runs.

\paragraph{Frontier chat models.}
GPT, Claude, Gemini and DeepSeek are asked, with one fixed English prompt and no tools, to return
an alpha-carbon-only PDB file for the sequence.
Because of cost they are run on a fixed, length-stratified subset of 180 targets (60 per set, at
most 300 residues), and every other system is rescored on the same subset for comparison.

\paragraph{Baseline checks.}
Appendix~\ref{app:repro} distinguishes paper-level reproductions, checks against earlier
runs, and baselines evaluated only under our protocol. Those distinctions apply to the
model comparisons throughout Section~\ref{sec:results}.

\section{Folding results}
\label{sec:results}

\subsection{Accuracy on three single-chain benchmarks}
\label{sec:floor}\label{sec:sets}
Under a fixed folding request, the final \calpha{} model uses Qwen residue states to
condition a frozen SimpleFold-3B head. It scores \N{ours_b2.foldbench334.overall},
\N{ours_b2.casp16.overall} and \N{ours_b2.cameo2025.overall} on FoldBench-334, CASP16
and CAMEO-2025, respectively. On all three sets it leads our joint and token variants,
while the public SimpleFold-3B model scores higher (Table~\ref{tab:main}, All columns).
FoldBench guided model development. On CASP16, nearest-structure retrieval scores
\N{retrieval.casp16.overall}; on CAMEO-2025 it reaches \N{retrieval.cameo2025.overall},
where 184 of 200 chains have a training homolog. Each mean therefore has a different
interpretation. The random-walk reference remains close to 0.14 on all three sets.

\begin{table}[htbp]
\centering
\caption{Mean TM-score by benchmark. In final \calpha{} (bold), a frozen SimpleFold-3B head predicts coordinates from Qwen residue states;
the FoldBench mean is the rightmost point in Figure~\ref{fig:ladder}. Failures score zero. The no-homolog columns cover targets
without a detected training homolog. Counts appear below the column heads.
$^\dagger$Uses an MSA. $^\ddagger$Published accuracy was not reproduced (BioMatrix) or
checked (MultiFlow). See Appendix~\ref{app:repro}.}
\label{tab:main}
\small
\begin{tabular*}{\linewidth}{@{\extracolsep{\fill}}lcccccc@{}}
\toprule
 & \multicolumn{2}{c}{FoldBench-334} & \multicolumn{2}{c}{CASP16} & \multicolumn{2}{c}{CAMEO-2025} \\
\cmidrule(lr){2-3}\cmidrule(lr){4-5}\cmidrule(lr){6-7}
Model & All & No hom. & All & No hom. & All & No hom. \\
 & \small(334) & \small(128) & \small(155) & \small(70) & \small(200) & \small(16) \\
\midrule
\multicolumn{7}{@{}l}{\emph{This work}} \\
\textbf{Calpha} (LLM + frozen 3B head) & \textbf{0.808} & \textbf{0.779} & \textbf{0.635} & \textbf{0.567} & \textbf{0.900} & \textbf{0.691} \\
Calpha-Joint (LLM + 100M head, co-trained) & 0.744 & 0.717 & 0.582 & 0.502 & 0.845 & 0.631 \\
Calpha-Token (DPLM-2 tokens) & 0.660 & 0.597 & 0.569 & 0.453 & 0.800 & 0.621 \\
\midrule
\multicolumn{7}{@{}l}{\emph{Folding models}} \\
ESMFold & 0.869 & 0.820 & 0.762 & 0.670 & 0.905 & 0.855 \\
SimpleFold-3B & 0.842 & 0.774 & 0.704 & 0.600 & 0.938 & 0.822 \\
SimpleFold-100M & 0.795 & 0.724 & 0.659 & 0.570 & 0.895 & 0.752 \\
OpenDDE (single sequence) & 0.438 & 0.503 & 0.368 & 0.366 & 0.370 & 0.398 \\
OpenDDE (+MSA)$^\dagger$ & 0.938 & 0.917 & 0.872 & 0.815 & 0.958 & 0.948 \\
\midrule
\multicolumn{7}{@{}l}{\emph{Sequence--structure models}} \\
ESM3-open (1.4B) & 0.787 & 0.720 & 0.643 & 0.543 & 0.852 & 0.739 \\
DPLM-2 (650M) & 0.656 & 0.582 & 0.515 & 0.428 & 0.757 & 0.608 \\
DPLM-2 Bit (650M) & 0.676 & 0.564 & 0.527 & 0.421 & 0.828 & 0.557 \\
MultiFlow$^\ddagger$ & 0.280 & 0.302 & 0.262 & 0.246 & 0.272 & 0.265 \\
\midrule
\multicolumn{7}{@{}l}{\emph{Sequence--structure--text LLMs}} \\
BioMatrix-1.7B$^\ddagger$ & 0.163 & 0.170 & 0.169 & 0.172 & 0.252 & 0.169 \\
BioMatrix-4B$^\ddagger$ & 0.158 & 0.164 & 0.163 & 0.161 & 0.311 & 0.145 \\
\bottomrule
\end{tabular*}

\end{table}

\subsection{Where the final systems differ}
\label{sec:toll}
ESM2 $\rightarrow$ SimpleFold-3B reaches \N{esm2_direct.foldbench334.overall}, \N{esm2_direct.casp16.overall}
and \N{esm2_direct.cameo2025.overall}. Both systems use the same pretrained head architecture,
but differ in the conditioning path, trainable weights, number of updates and final crop
(Table~\ref{tab:recipes}). Figure~\ref{fig:toll} compares their net difference in the two most populated
length ranges. Appendix Table~\ref{tab:toll} gives estimates, target counts and intervals for all groups.

\begin{figure}[!t]
\centering
\includegraphics[width=\linewidth]{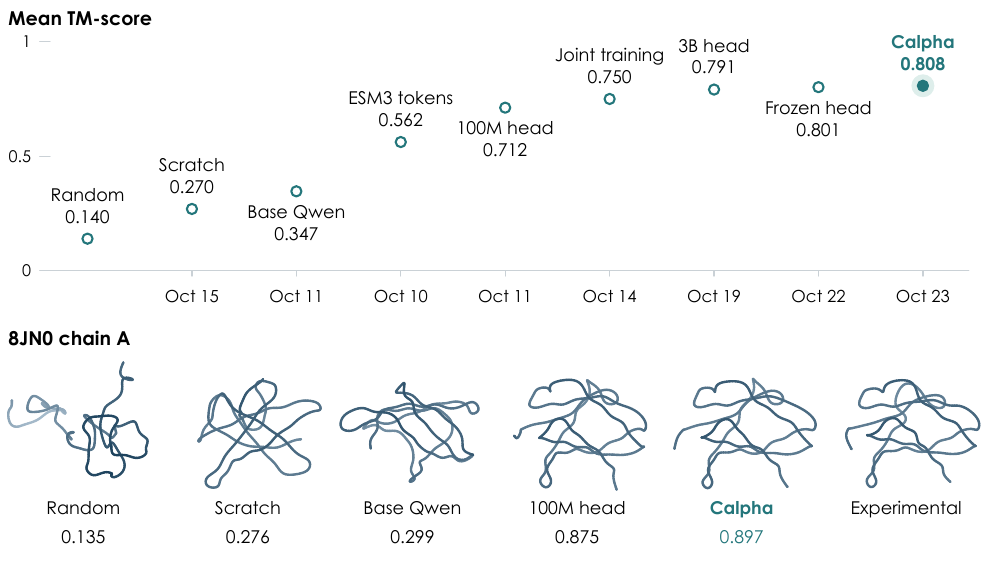}
\caption{FoldBench development runs, ordered by mean TM-score. The final \calpha{} point
matches Table~\ref{tab:main} (0.808). ESM3 tokens (0.562 on 333 chains) and Joint training
(0.750) are earlier runs. The table uses DPLM-2 tokens (0.660) and a later joint checkpoint
(0.744). All other plotted means cover 334 chains. Dates mark evaluation dates in October 2026. Below, each score is for 8JN0 chain A alone (89 residues), selected nearest
\calpha{}'s median, breaking ties by identifier. Traces share one view and scale.
Appendix~\ref{app:dev} specifies the training controls.}
\label{fig:ladder}
\end{figure}

At 100--299 residues the paired intervals include zero in all three sets. The CASP16 interval
extends to a loss of 0.030 TM, so differences of practical interest remain compatible with
these data. At 300--600 residues all three intervals fall below zero. The largest gap occurs
on CASP16, where \calpha{} outperforms the baseline on one of 34 entries. The 16 CASP16 domains
shorter than 100 residues also show a negative mean difference with an interval excluding zero.
Samples above 600 residues are small (12, 5 and 4 entries).

\begin{figure}[!b]
\centering
\includegraphics[width=0.78\linewidth]{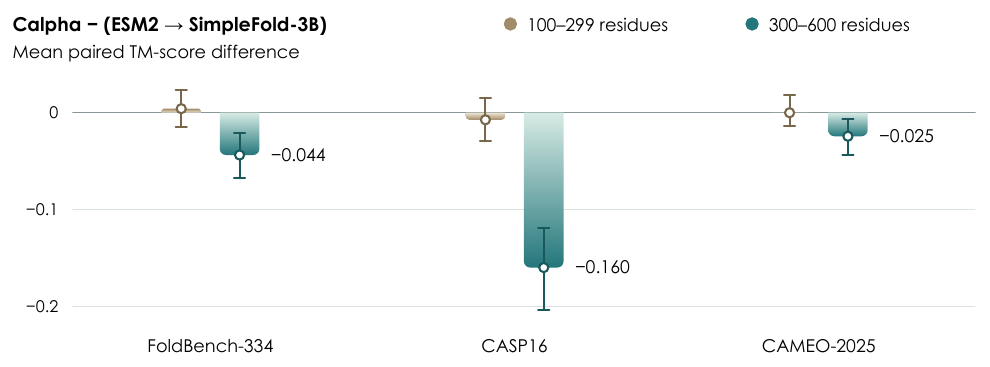}
\caption{The folding gap widens at 300--600 residues in each benchmark relative to 100--299 residues.
Bars show the mean within-target TM-score difference between Calpha and the ESM2 $\rightarrow$ SimpleFold-3B baseline.
Whiskers show bootstrap 95\% confidence intervals. Negative values favor the baseline.}
\label{fig:toll}
\end{figure}

\begin{figure}[!t]
\centering
\includegraphics[width=0.92\linewidth]{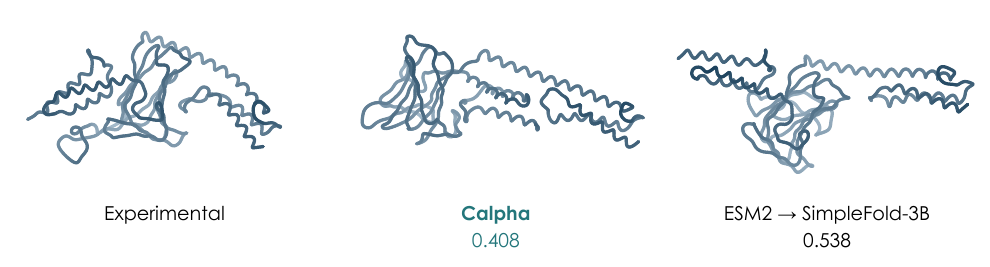}
\caption{CASP16 domain T1218-D2 (347 residues). Calpha scores 0.408 and the ESM2-conditioned
SimpleFold head scores 0.538. Their difference ($-0.130$) is closest to the group median
($-0.131$) among the 34 domains of length 300--600, with ties broken by identifier.
Backbones are aligned on common observed C$\alpha$ positions and share one view and scale.}
\label{fig:example}
\end{figure}

Figure~\ref{fig:example} shows a domain near the median of the long-chain comparison.
The homology split is less consistent across sets. The difference on the no-homolog bucket
is \N{ours_b2__vs__esm2_direct.foldbench334.no_family.delta} on FoldBench,
\N{ours_b2__vs__esm2_direct.casp16.no_family.delta} on CASP16 and
\N{ours_b2__vs__esm2_direct.cameo2025.no_family.delta} on CAMEO-2025. The last estimate
covers 16 entries, nine of which belong to three groups with over 95\% sequence identity.

\subsection{Comparison with specialist systems}
\label{sec:q2}
ESMFold exceeds \calpha{} on the development set and on CASP16. On CAMEO-2025 the
paired interval includes zero (Appendix~\ref{app:tables}). The public SimpleFold-3B
weights score higher on all three sets. Compared with ESM3-open, the paired intervals include
zero on FoldBench and CASP16; on CAMEO-2025 the interval favors \calpha{}. The public DPLM-2
models score lower under our protocol, although their paper's TM-scores on its own evaluation
set were not reproduced (Appendix~\ref{app:repro}). OpenDDE with an MSA is shown separately
to indicate the accuracy available with additional sequence information. Its sequence database
had no historical cutoff.

\subsection{Comparison with language-model outputs}
\label{sec:q1}
Calpha-Token (DPLM-2 tokens) reaches \N{ours_token.foldbench334.overall},
\N{ours_token.casp16.overall} and \N{ours_token.cameo2025.overall}. Its FoldBench mean
includes one missing prediction scored as zero across 334 targets. Figure~\ref{fig:ladder}
includes an earlier ESM3-token model because that was the backbone used for the first
flow-head experiment. BioMatrix's public checkpoints
perform close to the random-walk reference on FoldBench; their original folding results
were not reproduced. On CAMEO-2025, 15 sequences occur verbatim in the public BioMatrix
training set. Its 4B model scores 0.96 on these and 0.28 on the other 185.

On the common 180-target subset, four general-purpose chat models produced parseable
C$\alpha$ output on 707 of 720 requests. Their mean TM-scores lie between
\N{llm_deepseek.cameo2025.subset} and \N{llm_gpt.casp16.subset} under the fixed no-tool
prompt. Table~\ref{tab:llm} compares systems on the same sequences.

\begin{table}[!htb]
\centering
\caption{Frontier chat models asked to output an alpha-carbon-only PDB file, on a fixed subset of 60
targets per set ($\le$300 residues). TM-score, failures count as zero. Other systems are rescored
on the same targets.}
\label{tab:llm}
\small
\begin{tabular}{@{}lccc@{}}
\toprule
Model & FoldBench-334 & CASP16 & CAMEO-2025 \\
\midrule
\multicolumn{4}{@{}l}{\emph{General-purpose chat models (no tools)}} \\
GPT (gpt-6.1-sol) & 0.208 & 0.224 & 0.198 \\
Claude (claude-opus-5-5) & 0.213 & 0.196 & 0.196 \\
Gemini (gemini-4-argon) & 0.197 & 0.187 & 0.162 \\
DeepSeek (v4.1-flash) & 0.158 & 0.178 & 0.147 \\
\midrule
\multicolumn{4}{@{}l}{\emph{Same 180 targets, other systems}} \\
Random walk (floor) & 0.149 & 0.141 & 0.138 \\
BioMatrix-4B & 0.176 & 0.174 & 0.247 \\
ESM3-open & 0.760 & 0.659 & 0.844 \\
ESMFold & 0.841 & 0.750 & 0.902 \\
\textbf{Calpha} & \textbf{0.821} & \textbf{0.667} & \textbf{0.909} \\
\bottomrule
\end{tabular}

\end{table}

\section{Language behavior}
\label{sec:onemodel}

Section~\ref{sec:results} measured folding under a fixed request. In a conversation,
the user can instead ask a question about the same sequence, for which invoking the
structure head would be an error. We test whether Qwen makes that distinction under
different wordings. A separate before-and-after comparison measures how much general
question-answering accuracy it retains after protein training.

\paragraph{Deciding when to fold.}
For each eligible sequence we evaluate five folding requests and one protein question
that calls for a text answer. Table~\ref{tab:trigger} reports the rates. Four
request types elicited the trigger throughout the evaluated examples; the fifth elicited
it on 95.5--98.2\% across the sets. False triggers on the text questions ranged from
0 to 1.4\%. Appendix~\ref{app:repro} describes the length filter and the resulting counts.
An earlier model trained
on mixed Chinese--English data that also contained ``fold into structure tokens'' tasks
triggered on 5\% of plain English requests. With the architecture held fixed, the next run removed those token-folding tasks
and used English throughout. Triggering improved within the first quarter of training. The two
data changes were made together.

\begin{table}[t]
\centering
\caption{Fraction of requests answered with \texttt{<structure>} by the language model of
\calpha{}; one response per eligible target and phrasing.}
\label{tab:trigger}
\small
\begin{tabular}{@{}lccc@{}}
\toprule
Request & FoldBench-334 & CASP16 & CAMEO-2025 \\
\midrule
Training template & 1.000 & 1.000 & 1.000 \\
Unseen phrasing, close to training & 1.000 & 1.000 & 1.000 \\
Unseen phrasing, distant & 1.000 & 1.000 & 1.000 \\
After an unrelated first turn & 1.000 & 1.000 & 1.000 \\
Phrased like the language-side protein tasks & 0.982 & 0.964 & 0.955 \\
\midrule
Protein question that does not ask for a structure (should be 0) & 0.014 & 0.000 & 0.000 \\
\bottomrule
\end{tabular}
\end{table}

\paragraph{General ability.}
On a pre-registered 350-item subset of MMLU-Pro~\cite{wang2024mmlupro,mmlupro}, a harder successor of
MMLU~\cite{hendrycks2021mmlu}, with five-shot chain-of-thought
prompting, the base Qwen3.5-9B scores 0.774 and the stage-1 language model scores 0.703.
Accuracy fell by 7.1 percentage points. The trained model still produced answers in the
expected format. Reducing the share of general data from 25\% to 15\% cost a further five
points on the same test without improving folding.

\paragraph{From the trigger to a structure.}
A paraphrase can elicit the same trigger while changing the residue states read by the head.
The rates in Table~\ref{tab:trigger} therefore establish how often each request invokes folding. Measuring the
fraction of requests that actually return a full-length structure, and the TM-score of those
structures, requires following the same requests through coordinate generation and scoring.

\section{What the development experiments explain}
\label{sec:decomp}

Figure~\ref{fig:ladder} compares a random-walk reference with eight folding configurations.
Table~\ref{tab:diag} records which settings changed together. Each trained configuration
has one run (Appendix~\ref{app:dev}).

\begin{table}[!htb]
\centering
\caption{The comparisons behind each diagnostic conclusion (FoldBench-334, TM-score).}
\label{tab:diag}
\footnotesize
\setlength{\tabcolsep}{4pt}
\begin{tabular}{@{}p{3.3cm}p{2.9cm}p{3.6cm}p{1.9cm}p{3.6cm}@{}}
\toprule
Conclusion & Held fixed & Changed & Result & Also changed (confound) \\
\midrule
A 100M head improves on ESM3-token output &
frozen language model &
token output $\rightarrow$ flow-matching head (100M) on its hidden states &
0.562 $\rightarrow$ 0.712 &
pretrained structure weights; 333 vs.\ 334 scored \\
The protein-trained, ESM2-conditioned configuration gives higher TM in the 100M-head comparison &
head, recipe &
protein-trained residue states $\rightarrow$ states from sequence letters in untuned Qwen &
0.712 $\rightarrow$ 0.347 &
input feature conditions differ \\
Freezing the head worked best in this run &
language model, front-end, data &
frozen 3B head $\rightarrow$ unfrozen &
0.801 $\rightarrow$ 0.792 at best &
none known \\
Random initialization reduced TM at the measured step &
co-training recipe, data &
pretrained 100M head $\rightarrow$ random initialisation &
0.672 $\rightarrow$ 0.270 &
compared at one training step \\
A longer training crop helps long chains slightly &
starting checkpoint and data &
crop 256 $\rightarrow$ 512 residues &
$+0.006$ &
continued updates; one training run \\
\bottomrule
\end{tabular}
\end{table}

\paragraph{Output path.}
The early Qwen model writing ESM3 tokens scored 0.562. With a pretrained 100M flow
head reading the same frozen backbone, TM rose to 0.713 on the same 333 targets
(0.712 on all 334). This comparison includes the decoder's structural pretraining.
In the 100M-head experiments, one-hot sequence input and residue states from a Qwen
without protein training scored 0.346 and 0.347. A separate experiment used a trained
projector to inject ESM2 into a Qwen without protein training and obtained 0.476 with a
3B head. This shows that specialist features can pass through that backbone. The change
in head size and conditioning prevents attribution of the difference from 0.712 to
protein training alone.

\paragraph{Pretrained structural knowledge.}
The structure head was pretrained on far more structures than we train on.
With the head frozen and only the 8.4M-parameter front-end trained, \calpha{} reached 0.801;
unfreezing the head from that configuration made the model worse at every checkpoint we
measured, from 0.792 at best to 0.771 after 20{,}000 steps (Appendix~\ref{sec:freeze}).
A head initialised at random reached only 0.270 under our budget.
The historical comparisons with SimpleFold heads conditioned directly on ESM2 features gave gaps of 0.077 and 0.031 with the
100M and 3B heads on the development set. Their training configurations are described in
Appendix~\ref{sec:headsize}. The final 3B comparison changes the head training schedule
as well as the conditioning path.

\paragraph{What did not close the gap.}
Doubling the training steps, doubling the capacity of the front-end and letting gradients update
the language model each changed TM by less than 0.003.
Continuing training at crop 512 raised overall TM by 0.006 relative to the crop-256 run. The gain was concentrated in the 300--600-residue group.

\paragraph{Choosing among 16 samples.}
On CASP16, a single sample scores \N{ms.casp16.single}. Selecting the sample most similar
to the others gives \N{ms.casp16.medoid}. An oracle that sees the reference structure reaches
\N{ms.casp16.oracle}, close to the direct system's single-sample score of
\N{esm2_direct.casp16.overall}. The oracle bounds selection within these draws. Across
targets, agreement among samples correlates with true TM (Spearman
\N{ms.casp16.spearman}--\N{ms.cameo2025.spearman} across sets).
Using that agreement to assign calibrated confidence to a new target remains to be tested.

\begin{table}[!hb]
\centering
\caption{Sixteen samples per target from \calpha{} (TM-score). Every TM-score in the table is
computed against the true structure. ``Pick uses truth'' says whether \emph{choosing} the
sample requires the true structure: the oracle does, the consensus pick does not.}
\label{tab:multisample}
\small
\begin{tabular}{@{}lcccc@{}}
\toprule
Quantity & Pick uses truth & FoldBench-334 & CASP16 & CAMEO-2025 \\
\midrule
Single sample (as in Table~\ref{tab:main}) & no & 0.808 & 0.635 & 0.900 \\
Mean of 16 samples & -- & 0.808 & 0.633 & 0.902 \\
\addlinespace
Worst of 16 & yes & 0.773 & 0.589 & 0.886 \\
Best of 16 (oracle) & yes & 0.837 & 0.679 & 0.919 \\
Consensus pick (most similar to the others) & no & 0.814 & 0.642 & 0.903 \\
\addlinespace
Mean range of TM within a target & -- & 0.064 & 0.091 & 0.033 \\
Spearman: sample agreement vs.\ true TM & -- & 0.887 & 0.853 & 0.893 \\
\bottomrule
\end{tabular}

\end{table}

\section{Discussion, availability and outlook}
\label{sec:limits}
\subsection{Scope of the evidence}

The strongest length-dependent finding applies to the two trained systems in
Table~\ref{tab:recipes}. Their paired gap is negative at 300--600 residues on all three
sets. A matched comparison that fixes the trainable head parameters, optimizer schedule
and training crops across both conditioning paths would identify the contribution of
the path under that recipe. Explaining whether errors arise in local structure or in
the arrangement of distant regions also requires measurements across targets.

The reported scores assess C$\alpha$ structure for isolated chains or domains.
Table~\ref{tab:validity} reports the fraction of adjacent C$\alpha$ distances in the
physical range. A random walk constructed with physical step lengths also passes this
check. Side-chain placement, all-atom steric quality, complexes, inference latency
and peak inference memory remain unevaluated. Third-party training cutoffs and baseline
reproduction checks vary by model (Appendix~\ref{app:repro}).

Folding accuracy was measured with one request template. Request recognition under
paraphrases establishes a language-interface capability. It leaves open whether the
residue states produced by those phrasings generate structures of comparable accuracy.

\subsection{Relation to earlier systems}
\label{sec:related}

Single-sequence predictors such as ESMFold~\cite{lin2023esmfold} and OmegaFold~\cite{wu2022omegafold}
use protein language-model features to fold a sequence. SimpleFold~\cite{wang2025simplefold}
generates coordinates with a flow-matching transformer. Calpha uses its pretrained head
and supplies the conditioning features through Qwen.

Protein chat models such as ProtLLM~\cite{zhuo2024protllm} and ProteinGPT~\cite{xiao2024proteingpt}
answer questions in text. BioMatrix~\cite{pei2026biomatrix} and Trion-1~\cite{lu2026trion}
also generate discrete structure tokens. ESM3~\cite{hayes2025esm3} and DPLM-2~\cite{wang2025dplm2}
use discrete representations for joint sequence and structure modelling.
UniGenX~\cite{zhang2025unigenx} and HD-Prot~\cite{zhou2025hdprot} attach continuous
generators to sequence models. Calpha combines a chat interface with a continuous head:
the same language model decides when to fold and supplies the residue states used to
generate coordinates. A chat agent invoking an external folding program is another way
to handle these requests~\cite{ghafarollahi2024protagents,huang2025biomni}.
Comparing the two interfaces remains open.

\subsection{Availability and outlook}
\label{sec:road}
\begin{figure}[t]
\centering
\includegraphics[width=\linewidth]{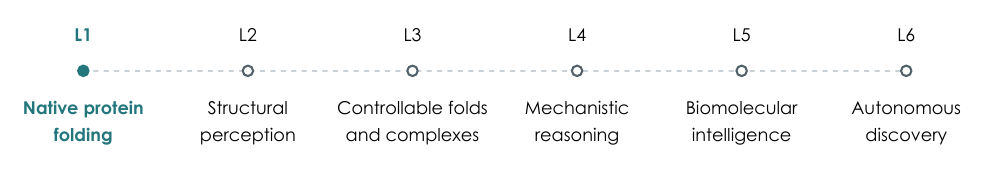}
\caption{The six levels of the \calpha{} roadmap span molecular generation and understanding.
This report evaluates sequence-to-structure generation at L1. Structural understanding at L1
and the later levels are research goals.}
\label{fig:roadmap}
\end{figure}

The code, the per-target scores and the scorer are available in the
\href{\calphagithub}{code repository}.
Figure~\ref{fig:roadmap} places single-chain folding within the project's longer-term goals.

\paragraph{Conclusion.}
Biomolecular foundation models and LLMs have so far been combined mainly through tool
calls or text-only interfaces. \calpha{} aims to make structure prediction and structure
understanding native abilities of a general-purpose LLM, so that reasoning in language
can shape what is predicted. This report takes the first step. Qwen learns when to invoke
folding across several English requests, and its residue states support single-chain
folding under the fixed benchmark request. We observe two limitations in these runs. After protein training,
accuracy is lower on the evaluated MMLU-Pro subset, and the final configuration scores lower than the ESM2
$\rightarrow$ SimpleFold-3B baseline on chains of 300--600 residues in all three sets.
It also leaves the central part of the goal untested. Routing through the language model
has not made folding more accurate, and we have not measured whether an instruction or a
line of reasoning changes the predicted coordinates. The next steps are to match the
training recipes so that the long-chain gap can be explained, to evaluate folding online
across phrasings, and to extend the model to complexes and to structure as an input
(Figure~\ref{fig:roadmap}).

\section*{Acknowledgements}
We thank the authors of SimpleFold~\cite{wang2025simplefold} for releasing their code and
pretrained weights. The structure head of \calpha{} and the direct ESM2 baseline are built
on them.

\bibliographystyle{oridynamic_odesign}
{\setlength{\bibsep}{3pt plus 0.3pt}\bibliography{references}}

\begin{thebibliography}{65}
\providecommand{\natexlab}[1]{#1}
\providecommand{\url}[1]{\texttt{#1}}
\expandafter\ifx\csname urlstyle\endcsname\relax
  \providecommand{\doi}[1]{doi: #1}\else
  \providecommand{\doi}{doi: \begingroup \urlstyle{rm}\Url}\fi

\bibitem[Anfinsen(1973)]{anfinsen1973}
Anfinsen, C.~B.
\newblock Principles that govern the folding of protein chains.
\newblock \emph{Science}, 181\penalty0 (4096):\penalty0 223--230, 1973.
\newblock \doi{10.1126/science.181.4096.223}.

\bibitem[Senior et~al.(2020)Senior, Evans, Jumper, et~al.]{senior2020alphafold}
Senior, A.~W., Evans, R., Jumper, J., et~al.
\newblock Improved protein structure prediction using potentials from deep
  learning.
\newblock \emph{Nature}, 577:\penalty0 706--710, 2020.

\bibitem[Jumper et~al.(2021)Jumper, Evans, Pritzel, et~al.]{jumper2021af2}
Jumper, J., Evans, R., Pritzel, A., et~al.
\newblock Highly accurate protein structure prediction with {AlphaFold}.
\newblock \emph{Nature}, 596:\penalty0 583--589, 2021.

\bibitem[Baek et~al.(2021)Baek, DiMaio, Anishchenko,
  et~al.]{baek2021rosettafold}
Baek, M., DiMaio, F., Anishchenko, I., et~al.
\newblock Accurate prediction of protein structures and interactions using a
  three-track neural network.
\newblock \emph{Science}, 373\penalty0 (6557):\penalty0 871--876, 2021.
\newblock \doi{10.1126/science.abj8754}.

\bibitem[Lin et~al.(2023)Lin, Akin, Rao, et~al.]{lin2023esmfold}
Lin, Z., Akin, H., Rao, R., et~al.
\newblock Evolutionary-scale prediction of atomic-level protein structure with
  a language model.
\newblock \emph{Science}, 379\penalty0 (6637):\penalty0 1123--1130, 2023.

\bibitem[Wu et~al.(2022)Wu, Ding, Wang, et~al.]{wu2022omegafold}
Wu, R., Ding, F., Wang, R., et~al.
\newblock High-resolution \emph{de novo} structure prediction from primary
  sequence.
\newblock bioRxiv preprint, 2022.

\bibitem[Rives et~al.(2021)Rives, Meier, Sercu, et~al.]{rives2021esm1b}
Rives, A., Meier, J., Sercu, T., et~al.
\newblock Biological structure and function emerge from scaling unsupervised
  learning to 250 million protein sequences.
\newblock \emph{Proceedings of the National Academy of Sciences}, 118\penalty0
  (15):\penalty0 e2016239118, 2021.
\newblock \doi{10.1073/pnas.2016239118}.

\bibitem[Elnaggar et~al.(2022)Elnaggar, Heinzinger, Dallago,
  et~al.]{elnaggar2022prottrans}
Elnaggar, A., Heinzinger, M., Dallago, C., et~al.
\newblock {ProtTrans}: Toward understanding the language of life through
  self-supervised learning.
\newblock \emph{IEEE Transactions on Pattern Analysis and Machine
  Intelligence}, 44\penalty0 (10):\penalty0 7112--7127, 2022.
\newblock \doi{10.1109/TPAMI.2021.3095381}.

\bibitem[Evans et~al.(2021)Evans, O'Neill, Pritzel,
  et~al.]{evans2021afmultimer}
Evans, R., O'Neill, M., Pritzel, A., et~al.
\newblock Protein complex prediction with {AlphaFold-Multimer}, 2021.
\newblock bioRxiv 10.1101/2021.10.04.463034.

\bibitem[Abramson et~al.(2024)Abramson, Adler, Dunger, et~al.]{abramson2024af3}
Abramson, J., Adler, J., Dunger, J., et~al.
\newblock Accurate structure prediction of biomolecular interactions with
  {AlphaFold} 3.
\newblock \emph{Nature}, 630:\penalty0 493--500, 2024.

\bibitem[Krishna et~al.(2024)Krishna, Wang, Ahern, et~al.]{krishna2024rfaa}
Krishna, R., Wang, J., Ahern, W., et~al.
\newblock Generalized biomolecular modeling and design with {RoseTTAFold
  All-Atom}.
\newblock \emph{Science}, 384\penalty0 (6693):\penalty0 eadl2528, 2024.

\bibitem[{Chai Discovery} et~al.(2024){Chai Discovery}, Boitreaud, Dent,
  McPartlon, Meier, Reis, Rogozhnikov, and Wu]{chai2024chai1}
{Chai Discovery}, Boitreaud, J., Dent, J., McPartlon, M., Meier, J., Reis, V.,
  Rogozhnikov, A., and Wu, K.
\newblock {Chai-1}: Decoding the molecular interactions of life.
\newblock bioRxiv preprint, 2024.

\bibitem[Wohlwend et~al.(2024)Wohlwend, Corso, Passaro,
  et~al.]{wohlwend2024boltz1}
Wohlwend, J., Corso, G., Passaro, S., et~al.
\newblock {Boltz-1}: Democratizing biomolecular interaction modeling.
\newblock bioRxiv preprint, 2024.

\bibitem[Passaro et~al.(2025)Passaro, Corso, Wohlwend,
  et~al.]{passaro2025boltz2}
Passaro, S., Corso, G., Wohlwend, J., et~al.
\newblock {Boltz-2}: Towards accurate and efficient binding affinity
  prediction.
\newblock \emph{bioRxiv preprint}
  \href{https://doi.org/10.1101/2025.06.14.659707}{2025.06.14.659707}, 2025.

\bibitem[{Aureka Research}(2026)]{opendde}
{Aureka Research}.
\newblock {OpenDDE}.
\newblock \url{https://github.com/aurekaresearch/OpenDDE}, 2026.
\newblock arXiv:2607.03787; \texttt{opendde==1.2.0}, accessed 2026-11-11.

\bibitem[Jing et~al.(2024)Jing, Berger, and Jaakkola]{jing2024alphaflow}
Jing, B., Berger, B., and Jaakkola, T.
\newblock {AlphaFold} meets flow matching for generating protein ensembles.
\newblock In \emph{Proceedings of the 41st International Conference on Machine
  Learning}, volume 235 of \emph{Proceedings of Machine Learning Research}, pp.
  22277--22303. PMLR, 2024.
\newblock \url{https://proceedings.mlr.press/v235/jing24a.html}.

\bibitem[Lipman et~al.(2023)Lipman, Chen, Ben-Hamu, Nickel, and
  Le]{lipman2023flow}
Lipman, Y., Chen, R. T.~Q., Ben-Hamu, H., Nickel, M., and Le, M.
\newblock Flow matching for generative modeling.
\newblock In \emph{International Conference on Learning Representations}, 2023.

\bibitem[Wang et~al.(2025{\natexlab{c}})Wang, Lu, Jaitly, Susskind, and
  Bautista]{wang2025simplefold}
Wang, Y., Lu, J., Jaitly, N., Susskind, J., and Bautista, M.~A.
\newblock {SimpleFold}: Folding proteins is simpler than you think,
  2025{\natexlab{c}}.
\newblock arXiv:2509.18480.

\bibitem[Watson et~al.(2023)Watson, Juergens, Bennett,
  et~al.]{watson2023rfdiffusion}
Watson, J.~L., Juergens, D., Bennett, N.~R., et~al.
\newblock De novo design of protein structure and function with {RFdiffusion}.
\newblock \emph{Nature}, 620\penalty0 (7976):\penalty0 1089--1100, 2023.
\newblock \doi{10.1038/s41586-023-06415-8}.

\bibitem[Ingraham et~al.(2023)Ingraham, Baranov, Costello,
  et~al.]{ingraham2023chroma}
Ingraham, J.~B., Baranov, M., Costello, Z., et~al.
\newblock Illuminating protein space with a programmable generative model.
\newblock \emph{Nature}, 623\penalty0 (7989):\penalty0 1070--1078, 2023.
\newblock \doi{10.1038/s41586-023-06728-8}.

\bibitem[Dauparas et~al.(2022)Dauparas, Anishchenko, Bennett,
  et~al.]{dauparas2022proteinmpnn}
Dauparas, J., Anishchenko, I., Bennett, N., et~al.
\newblock Robust deep learning--based protein sequence design using
  {ProteinMPNN}.
\newblock \emph{Science}, 378\penalty0 (6615):\penalty0 49--56, 2022.
\newblock \doi{10.1126/science.add2187}.

\bibitem[Hayes et~al.(2025)Hayes, Rao, Akin, et~al.]{hayes2025esm3}
Hayes, T., Rao, R., Akin, H., et~al.
\newblock Simulating 500 million years of evolution with a language model.
\newblock \emph{Science}, 387\penalty0 (6736):\penalty0 850--858, 2025.

\bibitem[Wang et~al.(2025{\natexlab{b}})Wang, Zheng, Ye, Xue, Huang, and
  Gu]{wang2025dplm2}
Wang, X., Zheng, Z., Ye, F., Xue, D., Huang, S., and Gu, Q.
\newblock {DPLM-2}: A multimodal diffusion protein language model.
\newblock In \emph{International Conference on Learning Representations},
  2025{\natexlab{b}}.

\bibitem[Vaswani et~al.(2017)Vaswani, Shazeer, Parmar, Uszkoreit, Jones, Gomez,
  Kaiser, and Polosukhin]{vaswani2017attention}
Vaswani, A., Shazeer, N., Parmar, N., Uszkoreit, J., Jones, L., Gomez, A.~N.,
  Kaiser, {\L}., and Polosukhin, I.
\newblock Attention is all you need.
\newblock In \emph{Advances in Neural Information Processing Systems}, 2017.

\bibitem[Brown et~al.(2020)Brown, Mann, Ryder, et~al.]{brown2020gpt3}
Brown, T.~B., Mann, B., Ryder, N., et~al.
\newblock Language models are few-shot learners.
\newblock In \emph{Advances in Neural Information Processing Systems}, 2020.

\bibitem[{OpenAI}(2023)]{openai2023gpt4}
{OpenAI}.
\newblock {GPT-4} technical report, 2023.
\newblock arXiv:2303.08774.

\bibitem[Yang et~al.(2025)Yang, Li, Yang, et~al.]{yang2025qwen3}
Yang, A., Li, A., Yang, B., et~al.
\newblock {Qwen3} technical report.
\newblock \emph{arXiv preprint} arXiv:2505.09388, 2025.

\bibitem[Wei et~al.(2022)Wei, Wang, Schuurmans, Bosma, Ichter, Xia, Chi, Le,
  and Zhou]{wei2022cot}
Wei, J., Wang, X., Schuurmans, D., Bosma, M., Ichter, B., Xia, F., Chi, E.~H.,
  Le, Q.~V., and Zhou, D.
\newblock Chain-of-thought prompting elicits reasoning in large language
  models.
\newblock In \emph{Advances in Neural Information Processing Systems}, 2022.

\bibitem[Guo et~al.(2025)Guo, Yang, Zhang, et~al.]{guo2025deepseekr1}
Guo, D., Yang, D., Zhang, H., et~al.
\newblock {DeepSeek-R1}: Incentivizing reasoning capability in {LLMs} via
  reinforcement learning, 2025.
\newblock arXiv:2501.12948.

\bibitem[Yao et~al.(2023)Yao, Zhao, Yu, Du, Shafran, Narasimhan, and
  Cao]{yao2023react}
Yao, S., Zhao, J., Yu, D., Du, N., Shafran, I., Narasimhan, K., and Cao, Y.
\newblock {ReAct}: Synergizing reasoning and acting in language models.
\newblock In \emph{International Conference on Learning Representations}, 2023.

\bibitem[Boiko et~al.(2023)Boiko, MacKnight, Kline, and
  Gomes]{boiko2023coscientist}
Boiko, D.~A., MacKnight, R., Kline, B., and Gomes, G.
\newblock Autonomous chemical research with large language models.
\newblock \emph{Nature}, 624\penalty0 (7992):\penalty0 570--578, 2023.
\newblock \doi{10.1038/s41586-023-06792-0}.

\bibitem[{M. Bran} et~al.(2024){M. Bran}, Cox, Schilter, Baldassari, White, and
  Schwaller]{bran2024chemcrow}
{M. Bran}, A., Cox, S., Schilter, O., Baldassari, C., White, A.~D., and
  Schwaller, P.
\newblock Augmenting large language models with chemistry tools.
\newblock \emph{Nature Machine Intelligence}, 6\penalty0 (5):\penalty0
  525--535, 2024.
\newblock \doi{10.1038/s42256-024-00832-8}.

\bibitem[Gottweis et~al.(2025)Gottweis, Weng, Daryin,
  et~al.]{gottweis2025coscientist}
Gottweis, J., Weng, W.-H., Daryin, A., et~al.
\newblock Towards an {AI} co-scientist, 2025.
\newblock arXiv:2502.18864.

\bibitem[Swanson et~al.(2025)Swanson, Wu, Bulaong, Pak, and
  Zou]{swanson2025virtuallab}
Swanson, K., Wu, W., Bulaong, N.~L., Pak, J.~E., and Zou, J.
\newblock The virtual lab of {AI} agents designs new {SARS-CoV-2} nanobodies.
\newblock \emph{Nature}, 646\penalty0 (8085):\penalty0 716--723, 2025.
\newblock \doi{10.1038/s41586-025-09442-9}.

\bibitem[Huang et~al.(2025)Huang, Zhang, Wang, et~al.]{huang2025biomni}
Huang, K., Zhang, S., Wang, H., et~al.
\newblock Biomni: A general-purpose biomedical {AI} agent.
\newblock \emph{bioRxiv preprint}
  \href{https://doi.org/10.1101/2025.05.30.656746}{2025.05.30.656746}, 2025.

\bibitem[Ghafarollahi and Buehler(2024)]{ghafarollahi2024protagents}
Ghafarollahi, A. and Buehler, M.~J.
\newblock {ProtAgents}: protein discovery via large language model multi-agent
  collaborations combining physics and machine learning.
\newblock \emph{Digital Discovery}, 3\penalty0 (7):\penalty0 1389--1409, 2024.
\newblock \doi{10.1039/D4DD00013G}.

\bibitem[Ge et~al.(2026)Ge, Zhu, Zhang, et~al.]{ge2026autobinder}
Ge, F., Zhu, J., Zhang, L., et~al.
\newblock {AutoBinder Agent}: An {MCP}-based agent for end-to-end protein
  binder design.
\newblock \emph{arXiv preprint arXiv:2602.00019}, 2026.

\bibitem[Xu et~al.(2023)Xu, Yuan, Miret, and Tang]{xu2023protst}
Xu, M., Yuan, X., Miret, S., and Tang, J.
\newblock {ProtST}: Multi-modality learning of protein sequences and biomedical
  texts.
\newblock In \emph{Proceedings of the 40th International Conference on Machine
  Learning}, volume 202 of \emph{Proceedings of Machine Learning Research}, pp.
  38749--38767. PMLR, 2023.
\newblock \url{https://proceedings.mlr.press/v202/xu23t.html}.

\bibitem[Zhuo et~al.(2024)Zhuo, Chi, Xu, et~al.]{zhuo2024protllm}
Zhuo, L., Chi, Z., Xu, M., et~al.
\newblock {ProtLLM}: An interleaved protein-language {LLM} with protein-as-word
  pre-training.
\newblock In \emph{Proceedings of the 62nd Annual Meeting of the Association
  for Computational Linguistics (Volume 1: Long Papers)}, pp. 8950--8963.
  Association for Computational Linguistics, 2024.
\newblock \doi{10.18653/v1/2024.acl-long.484}.

\bibitem[Xiao et~al.(2024)Xiao, Sun, Jin, Wang, and Wang]{xiao2024proteingpt}
Xiao, Y., Sun, E., Jin, Y., Wang, Q., and Wang, W.
\newblock {ProteinGPT}: Multimodal {LLM} for protein property prediction and
  structure understanding.
\newblock \emph{arXiv preprint} arXiv:2408.11363, 2024.
\newblock Spotlight at the Machine Learning for Genomics Explorations workshop,
  ICLR 2025.

\bibitem[Wang et~al.(2025{\natexlab{a}})Wang, Fan, Quan, Yao, and
  Yang]{wang2024protchatgpt}
Wang, C., Fan, H., Quan, R., Yao, L., and Yang, Y.
\newblock {ProtChatGPT}: Towards understanding proteins with hybrid
  representation and large language models.
\newblock In \emph{Proceedings of the 48th International ACM SIGIR Conference
  on Research and Development in Information Retrieval}, pp. 1076--1086. ACM,
  2025{\natexlab{a}}.
\newblock \doi{10.1145/3726302.3730064}.

\bibitem[Zhou et~al.(2025{\natexlab{b}})Zhou, Han, Zhang,
  et~al.]{zhou2025evolla}
Zhou, X., Han, C., Zhang, Y., et~al.
\newblock Decoding the molecular language of proteins with {Evolla}.
\newblock bioRxiv preprint, 2025{\natexlab{b}}.

\bibitem[Alayrac et~al.(2022)Alayrac, Donahue, Luc,
  et~al.]{alayrac2022flamingo}
Alayrac, J.-B., Donahue, J., Luc, P., et~al.
\newblock Flamingo: a visual language model for few-shot learning.
\newblock In \emph{Advances in Neural Information Processing Systems},
  volume~35, pp. 23716--23736. Curran Associates, Inc., 2022.

\bibitem[Li et~al.(2023)Li, Li, Savarese, and Hoi]{li2023blip2}
Li, J., Li, D., Savarese, S., and Hoi, S.
\newblock {BLIP-2}: Bootstrapping language-image pre-training with frozen image
  encoders and large language models.
\newblock In \emph{Proceedings of the 40th International Conference on Machine
  Learning}, volume 202 of \emph{Proceedings of Machine Learning Research}, pp.
  19730--19742. PMLR, 2023.
\newblock \url{https://proceedings.mlr.press/v202/li23q.html}.

\bibitem[Liu et~al.(2023)Liu, Li, Wu, and Lee]{liu2023llava}
Liu, H., Li, C., Wu, Q., and Lee, Y.~J.
\newblock Visual instruction tuning.
\newblock In \emph{Advances in Neural Information Processing Systems},
  volume~36, pp. 34892--34916. Curran Associates, Inc., 2023.

\bibitem[Chen et~al.(2024)Chen, Chen, Xie, Xue, Zhang, Zhou, and
  Fang]{chen2024helixprotx}
Chen, Z., Chen, T., Xie, C., Xue, Y., Zhang, X., Zhou, J., and Fang, X.
\newblock Unifying sequences, structures, and descriptions for any-to-any
  protein generation with the large multimodal model {HelixProtX}.
\newblock \emph{arXiv preprint arXiv:2407.09274}, 2024.

\bibitem[Ma et~al.(2025)Ma, Fan, Wang, et~al.]{ma2025prottex}
Ma, Z., Fan, C., Wang, Z., et~al.
\newblock {ProtTeX}: Structure-in-context reasoning and editing of proteins
  with large language models.
\newblock \emph{arXiv preprint arXiv:2503.08179}, 2025.

\bibitem[Pei et~al.(2026)Pei, Zhou, Duan, Zhao, Li, Guo, He, Li, Hsieh, He,
  Yan, and Wu]{pei2026biomatrix}
Pei, Q., Zhou, Z., Duan, Y., Zhao, Y., Li, W., Guo, H., He, L., Li, C., Hsieh,
  C.-Y., He, C., Yan, R., and Wu, L.
\newblock {BioMatrix}: Towards a comprehensive biological foundation model
  spanning the modality matrix of sequences, structures, and language.
\newblock \emph{arXiv preprint} arXiv:2606.22138, 2026.
\newblock arXiv:2606.22138.

\bibitem[Lu et~al.(2026)Lu, Yang, Feng, Qiu, Ouyang, Wu, Zhang, Wang, Ma, and
  Zhou]{lu2026trion}
Lu, T., Yang, K., Feng, J., Qiu, K., Ouyang, Y., Wu, Y., Zhang, X., Wang, L.,
  Ma, W.-Y., and Zhou, H.
\newblock {Trion-1}: Unifying language, protein sequence, and structure with
  diffusion language models.
\newblock \emph{arXiv preprint} arXiv:2610.31215, 2026.
\newblock arXiv:2610.31215.

\bibitem[Zhang et~al.(2025)Zhang, Li, Luo, et~al.]{zhang2025unigenx}
Zhang, G., Li, Y., Luo, R., et~al.
\newblock {UniGenX}: a unified generative foundation model that couples
  sequence, structure and function to accelerate scientific design across
  proteins, molecules and materials.
\newblock \emph{arXiv preprint arXiv:2503.06687}, 2025.

\bibitem[Zhou et~al.(2025{\natexlab{c}})Zhou, Qu, Liu, Lin, Song, and
  Fan]{zhou2025hdprot}
Zhou, Y., Qu, H., Liu, Y., Lin, S., Song, L., and Fan, W.
\newblock {HD-Prot}: A protein language model for joint sequence-structure
  modeling with continuous structure tokens.
\newblock \emph{arXiv preprint arXiv:2512.15133}, 2025{\natexlab{c}}.
\newblock Long version of a paper to appear at KDD 2026.

\bibitem[Li et~al.(2024)Li, Tian, Li, Deng, and He]{li2024mar}
Li, T., Tian, Y., Li, H., Deng, M., and He, K.
\newblock Autoregressive image generation without vector quantization.
\newblock In \emph{Advances in Neural Information Processing Systems},
  volume~37, pp. 56424--56445. Curran Associates, Inc., 2024.

\bibitem[Zhou et~al.(2025{\natexlab{a}})Zhou, Yu, Babu,
  et~al.]{zhou2024transfusion}
Zhou, C., Yu, L., Babu, A., et~al.
\newblock Transfusion: Predict the next token and diffuse images with one
  multi-modal model.
\newblock In \emph{International Conference on Learning Representations
  (ICLR)}, 2025{\natexlab{a}}.
\newblock \url{https://openreview.net/forum?id=SI2hI0frk6}.

\bibitem[Black et~al.(2025)Black, Brown, Driess, et~al.]{black2024pi0}
Black, K., Brown, N., Driess, D., et~al.
\newblock {$\pi_0$}: A vision-language-action flow model for general robot
  control.
\newblock In \emph{Robotics: Science and Systems XXI}, 2025.
\newblock \doi{10.15607/RSS.2025.XXI.010}.

\bibitem[Qwen()]{qwen35}
Qwen.
\newblock {Qwen3.5-9B}.
\newblock Model card, Alibaba Cloud.
\newblock Accessed 2026.

\bibitem[Wang et~al.(2024)Wang, Ma, Zhang, et~al.]{wang2024mmlupro}
Wang, Y., Ma, X., Zhang, G., et~al.
\newblock {MMLU-Pro}: A more robust and challenging multi-task language
  understanding benchmark.
\newblock In \emph{Advances in Neural Information Processing Systems (Datasets
  and Benchmarks Track)}, volume~37, pp. 95266--95290. Curran Associates, Inc.,
  2024.

\bibitem[Liu(2026)]{companion}
Liu, Y.
\newblock Writability: What makes a protein structure tokenizer a good vehicle
  for a language model, 2026.
\newblock Companion technical report.

\bibitem[Robin et~al.(2021)Robin, Haas, Gumienny, Smolinski, Tauriello, and
  Schwede]{robin2021cameo}
Robin, X., Haas, J., Gumienny, R., Smolinski, A., Tauriello, G., and Schwede,
  T.
\newblock {Continuous Automated Model EvaluatiOn (CAMEO)}---perspectives on the
  future of fully automated evaluation of structure prediction methods.
\newblock \emph{Proteins: Structure, Function, and Bioinformatics}, 89\penalty0
  (12):\penalty0 1977--1986, 2021.
\newblock \doi{10.1002/prot.26213}.

\bibitem[Berman et~al.(2000)Berman, Westbrook, Feng, Gilliland, Bhat, Weissig,
  Shindyalov, and Bourne]{berman2000pdb}
Berman, H.~M., Westbrook, J., Feng, Z., Gilliland, G., Bhat, T.~N., Weissig,
  H., Shindyalov, I.~N., and Bourne, P.~E.
\newblock The {Protein Data Bank}.
\newblock \emph{Nucleic Acids Research}, 28\penalty0 (1):\penalty0 235--242,
  2000.
\newblock \doi{10.1093/nar/28.1.235}.

\bibitem[Varadi et~al.(2022)Varadi, Anyango, Deshpande, et~al.]{varadi2022afdb}
Varadi, M., Anyango, S., Deshpande, M., et~al.
\newblock {AlphaFold Protein Structure Database}: massively expanding the
  structural coverage of protein-sequence space with high-accuracy models.
\newblock \emph{Nucleic Acids Research}, 50\penalty0 (D1):\penalty0 D439--D444,
  2022.
\newblock \doi{10.1093/nar/gkab1061}.

\bibitem[Zhang and Skolnick(2004)]{tmscore}
Zhang, Y. and Skolnick, J.
\newblock Scoring function for automated assessment of protein structure
  template quality.
\newblock \emph{Proteins: Structure, Function, and Bioinformatics}, 57\penalty0
  (4):\penalty0 702--710, 2004.

\bibitem[Mariani et~al.(2013)Mariani, Biasini, Barbato, and Schwede]{lddt}
Mariani, V., Biasini, M., Barbato, A., and Schwede, T.
\newblock lddt: a local superposition-free score for comparing protein
  structures and models using distance difference tests.
\newblock \emph{Bioinformatics}, 29\penalty0 (21):\penalty0 2722--2728, 2013.

\bibitem[MMLU-Pro()]{mmlupro}
MMLU-Pro.
\newblock {MMLU-Pro}: A more robust and challenging multi-task language
  understanding benchmark.
\newblock \url{https://huggingface.co/datasets/TIGER-Lab/MMLU-Pro}.
\newblock Evaluated on a pre-registered 350-item stratified subset with the
  upstream item list and scorer.

\bibitem[Hendrycks et~al.(2021)Hendrycks, Burns, Basart, Zou, Mazeika, Song,
  and Steinhardt]{hendrycks2021mmlu}
Hendrycks, D., Burns, C., Basart, S., Zou, A., Mazeika, M., Song, D., and
  Steinhardt, J.
\newblock Measuring massive multitask language understanding.
\newblock In \emph{International Conference on Learning Representations
  (ICLR)}, 2021.

\bibitem[Mirdita et~al.(2022)Mirdita, Sch{\"u}tze, Moriwaki, Heo, Ovchinnikov,
  and Steinegger]{mirdita2022colabfold}
Mirdita, M., Sch{\"u}tze, K., Moriwaki, Y., Heo, L., Ovchinnikov, S., and
  Steinegger, M.
\newblock {ColabFold}: making protein folding accessible to all.
\newblock \emph{Nature Methods}, 19\penalty0 (6):\penalty0 679--682, 2022.
\newblock \doi{10.1038/s41592-022-01488-1}.

\end{thebibliography}

\clearpage

\appendix
{\titlefont\sffamily\seedblue{Appendix}\par}
\vskip 4mm

\section{Additional Tables}
\label{app:tables}

\begin{table}[H]
\centering
\caption{System comparison: paired $\Delta$TM of \calpha{} minus the ESM2 $\rightarrow$ SimpleFold-3B baseline (training configurations in Table~\ref{tab:recipes}). $^{*}$: bootstrap 95\% CI excludes
zero. CAMEO-2025 has four targets above 600 residues, too few for an interval.}
\label{tab:toll}
\footnotesize
\setlength{\tabcolsep}{4pt}
\begin{tabular}{@{}lrlrlrl@{}}
\toprule
 & \multicolumn{2}{c}{FoldBench-334} & \multicolumn{2}{c}{CASP16} & \multicolumn{2}{c}{CAMEO-2025} \\
\cmidrule(lr){2-3}\cmidrule(lr){4-5}\cmidrule(lr){6-7}
Bucket & $n$ & $\Delta$TM [95\% CI] & $n$ & $\Delta$TM [95\% CI] & $n$ & $\Delta$TM [95\% CI] \\
\midrule
All & 334 & $-0.014$ {\scriptsize[$-0.028$, $+0.000$]} & 155 & $-0.047$ {\scriptsize[$-0.068$, $-0.026$]}$^{*}$ & 200 & $-0.017$ {\scriptsize[$-0.030$, $-0.004$]}$^{*}$ \\
\addlinespace
Homolog & 206 & $-0.024$ {\scriptsize[$-0.039$, $-0.009$]}$^{*}$ & 85 & $-0.068$ {\scriptsize[$-0.098$, $-0.039$]}$^{*}$ & 184 & $-0.010$ {\scriptsize[$-0.023$, $+0.003$]} \\
No homolog & 128 & $+0.002$ {\scriptsize[$-0.026$, $+0.032$]} & 70 & $-0.022$ {\scriptsize[$-0.052$, $+0.007$]} & 16 & $-0.100$ {\scriptsize[$-0.143$, $-0.060$]}$^{*}$ \\
\addlinespace
$<$100 & 38 & $-0.013$ {\scriptsize[$-0.064$, $+0.043$]} & 16 & $-0.024$ {\scriptsize[$-0.051$, $-0.003$]}$^{*}$ & 7 & $-0.017$ {\scriptsize[$-0.070$, $+0.018$]} \\
100--299 & 188 & $+0.004$ {\scriptsize[$-0.015$, $+0.023$]} & 100 & $-0.008$ {\scriptsize[$-0.030$, $+0.015$]} & 101 & $-0.000$ {\scriptsize[$-0.014$, $+0.018$]} \\
300--600 & 96 & $-0.044$ {\scriptsize[$-0.068$, $-0.021$]}$^{*}$ & 34 & $-0.160$ {\scriptsize[$-0.204$, $-0.119$]}$^{*}$ & 88 & $-0.025$ {\scriptsize[$-0.044$, $-0.007$]}$^{*}$ \\
$>$600 & 12 & $-0.062$ {\scriptsize[$-0.146$, $+0.019$]} & 5 & $-0.143$ {\scriptsize[$-0.375$, $-0.019$]}$^{*}$ & -- & -- \\
\bottomrule
\end{tabular}

\end{table}

\begin{table}[H]
\centering
\caption{Completion rate (``Done'') and fraction of consecutive alpha-carbon distances in
3.6--4.0\,\AA{} (``Valid''). The check detects broken or stretched chains only: the random
walk passes it by construction. The first row is the ground truth passed through the same
code. A dash marks a metric that was not recorded.}
\label{tab:validity}
\small
\begin{tabular}{@{}lcccccc@{}}
\toprule
 & \multicolumn{2}{c}{FoldBench-334} & \multicolumn{2}{c}{CASP16} & \multicolumn{2}{c}{CAMEO-2025} \\
\cmidrule(lr){2-3}\cmidrule(lr){4-5}\cmidrule(lr){6-7}
Model & Done & Valid & Done & Valid & Done & Valid \\
\midrule
Ground truth (ceiling) & 1.00 & 0.996 & 1.00 & 0.996 & 1.00 & 0.996 \\
\textbf{Calpha} (LLM + frozen 3B head) & 1.00 & 0.995 & 1.00 & 0.997 & 1.00 & 0.998 \\
Calpha-Joint (LLM + 100M head, co-trained) & 1.00 & 0.943 & 1.00 & 0.942 & 1.00 & 0.962 \\
Calpha-Token (DPLM-2 tokens) & 1.00 & -- & 1.00 & -- & 1.00 & -- \\
ESM2 $\rightarrow$ SimpleFold-3B (fine-tuned) & 1.00 & 0.993 & 1.00 & 0.996 & 1.00 & 0.992 \\
ESMFold & 1.00 & 0.994 & 1.00 & 0.993 & 1.00 & 0.996 \\
SimpleFold-3B & 1.00 & 0.998 & 1.00 & 0.998 & 1.00 & 0.998 \\
SimpleFold-100M & 1.00 & 0.997 & 1.00 & 0.997 & 1.00 & 0.998 \\
OpenDDE (single sequence) & 1.00 & 0.999 & 1.00 & 0.999 & 1.00 & 0.999 \\
OpenDDE (+MSA)$^\dagger$ & 1.00 & 0.998 & 1.00 & 0.999 & 1.00 & 0.998 \\
ESM3-open (1.4B) & 1.00 & 0.871 & 1.00 & 0.791 & 1.00 & 0.888 \\
DPLM-2 (650M) & 1.00 & 0.969 & 1.00 & 0.961 & 1.00 & 0.976 \\
DPLM-2 Bit (650M) & 1.00 & 0.978 & 1.00 & 0.965 & 1.00 & 0.988 \\
MultiFlow$^\ddagger$ & 1.00 & 0.980 & 1.00 & 0.979 & 1.00 & 0.985 \\
BioMatrix-1.7B$^\ddagger$ & 1.00 & 0.885 & 1.00 & 0.910 & 1.00 & 0.885 \\
BioMatrix-4B$^\ddagger$ & 1.00 & 0.907 & 1.00 & 0.915 & 1.00 & 0.908 \\
Nearest training structure (copy) & 0.90 & 0.986 & 0.82 & 0.991 & 0.98 & 0.993 \\
Random walk (floor) & 1.00 & 1.000 & 1.00 & 1.000 & 1.00 & 1.000 \\
\bottomrule
\end{tabular}

\end{table}

\begin{table}[H]
\centering
\caption{Paired $\Delta$TM of \calpha{} minus each other system over all targets of a set, with
bootstrap 95\% confidence intervals. $^{*}$: the interval excludes zero. Positive values mean
\calpha{} is better.}
\label{tab:paired}
\footnotesize
\begin{tabular}{@{}llll@{}}
\toprule
\calpha{} minus & FoldBench-334 & CASP16 & CAMEO-2025 \\
\midrule
Calpha-Joint & $+0.063$ {\scriptsize[$+0.052$, $+0.075$]}$^{*}$ & $+0.052$ {\scriptsize[$+0.038$, $+0.068$]}$^{*}$ & $+0.055$ {\scriptsize[$+0.043$, $+0.068$]}$^{*}$ \\
Calpha-Token & $+0.147$ {\scriptsize[$+0.130$, $+0.165$]}$^{*}$ & $+0.066$ {\scriptsize[$+0.048$, $+0.084$]}$^{*}$ & $+0.100$ {\scriptsize[$+0.081$, $+0.120$]}$^{*}$ \\
ESM2 $\rightarrow$ SimpleFold-3B (fine-tuned) & $-0.014$ {\scriptsize[$-0.028$, $+0.000$]} & $-0.047$ {\scriptsize[$-0.068$, $-0.026$]}$^{*}$ & $-0.017$ {\scriptsize[$-0.030$, $-0.004$]}$^{*}$ \\
ESMFold & $-0.062$ {\scriptsize[$-0.080$, $-0.043$]}$^{*}$ & $-0.128$ {\scriptsize[$-0.154$, $-0.102$]}$^{*}$ & $-0.005$ {\scriptsize[$-0.028$, $+0.019$]} \\
SimpleFold-3B & $-0.034$ {\scriptsize[$-0.052$, $-0.016$]}$^{*}$ & $-0.070$ {\scriptsize[$-0.093$, $-0.047$]}$^{*}$ & $-0.038$ {\scriptsize[$-0.052$, $-0.023$]}$^{*}$ \\
SimpleFold-100M & $+0.013$ {\scriptsize[$-0.004$, $+0.030$]} & $-0.025$ {\scriptsize[$-0.043$, $-0.007$]}$^{*}$ & $+0.005$ {\scriptsize[$-0.009$, $+0.021$]} \\
OpenDDE (single sequence) & $+0.370$ {\scriptsize[$+0.340$, $+0.399$]}$^{*}$ & $+0.266$ {\scriptsize[$+0.229$, $+0.305$]}$^{*}$ & $+0.530$ {\scriptsize[$+0.501$, $+0.559$]}$^{*}$ \\
ESM3-open (1.4B) & $+0.021$ {\scriptsize[$-0.000$, $+0.042$]} & $-0.008$ {\scriptsize[$-0.031$, $+0.013$]} & $+0.048$ {\scriptsize[$+0.023$, $+0.074$]}$^{*}$ \\
DPLM-2 (650M) & $+0.151$ {\scriptsize[$+0.133$, $+0.170$]}$^{*}$ & $+0.120$ {\scriptsize[$+0.100$, $+0.141$]}$^{*}$ & $+0.143$ {\scriptsize[$+0.118$, $+0.170$]}$^{*}$ \\
DPLM-2 Bit (650M) & $+0.132$ {\scriptsize[$+0.111$, $+0.152$]}$^{*}$ & $+0.108$ {\scriptsize[$+0.082$, $+0.133$]}$^{*}$ & $+0.072$ {\scriptsize[$+0.052$, $+0.093$]}$^{*}$ \\
BioMatrix-4B$^\ddagger$ & $+0.649$ {\scriptsize[$+0.626$, $+0.671$]}$^{*}$ & $+0.471$ {\scriptsize[$+0.433$, $+0.510$]}$^{*}$ & $+0.589$ {\scriptsize[$+0.546$, $+0.630$]}$^{*}$ \\
Nearest training structure (copy) & $+0.261$ {\scriptsize[$+0.218$, $+0.304$]}$^{*}$ & $+0.165$ {\scriptsize[$+0.101$, $+0.229$]}$^{*}$ & $+0.056$ {\scriptsize[$+0.028$, $+0.085$]}$^{*}$ \\
\bottomrule
\end{tabular}

\end{table}

\begin{table}[H]
\centering
\caption{TM-score by chain length, FoldBench-334. Target counts under the column heads.}
\label{tab:len-fb}
\footnotesize\renewcommand{\arraystretch}{0.93}
\begin{tabular}{@{}lcccc@{}}
\toprule
Model & $<$100 & 100--299 & 300--600 & $>$600 \\
 & \small(38) & \small(188) & \small(96) & \small(12) \\
\midrule
\textbf{Calpha} (LLM + frozen 3B head) & \textbf{0.821} & \textbf{0.817} & \textbf{0.817} & \textbf{0.536} \\
Calpha-Joint (LLM + 100M head, co-trained) & 0.805 & 0.779 & 0.699 & 0.369 \\
Calpha-Token (DPLM-2 tokens) & 0.650 & 0.653 & 0.708 & 0.423 \\
ESM2 $\rightarrow$ SimpleFold-3B (fine-tuned) & 0.833 & 0.813 & 0.861 & 0.598 \\
ESMFold & 0.801 & 0.853 & 0.930 & 0.855 \\
SimpleFold-3B & 0.756 & 0.823 & 0.914 & 0.821 \\
SimpleFold-100M & 0.737 & 0.774 & 0.866 & 0.741 \\
OpenDDE (single sequence) & 0.593 & 0.454 & 0.360 & 0.307 \\
OpenDDE (+MSA)$^\dagger$ & 0.894 & 0.931 & 0.967 & 0.967 \\
ESM3-open (1.4B) & 0.708 & 0.757 & 0.875 & 0.805 \\
DPLM-2 (650M) & 0.621 & 0.635 & 0.731 & 0.510 \\
DPLM-2 Bit (650M) & 0.598 & 0.653 & 0.760 & 0.616 \\
MultiFlow$^\ddagger$ & 0.327 & 0.292 & 0.250 & 0.181 \\
BioMatrix-1.7B$^\ddagger$ & 0.195 & 0.168 & 0.146 & 0.119 \\
BioMatrix-4B$^\ddagger$ & 0.187 & 0.168 & 0.135 & 0.101 \\
Nearest training structure (copy) & 0.436 & 0.484 & 0.712 & 0.559 \\
Random walk (floor) & 0.152 & 0.143 & 0.133 & 0.126 \\
\bottomrule
\end{tabular}

\end{table}

\begin{table}[H]
\centering
\caption{TM-score by chain length, CASP16.}
\label{tab:len-casp}
\footnotesize\renewcommand{\arraystretch}{0.93}
\begin{tabular}{@{}lcccc@{}}
\toprule
Model & $<$100 & 100--299 & 300--600 & $>$600 \\
 & \small(16) & \small(100) & \small(34) & \small(5) \\
\midrule
\textbf{Calpha} (LLM + frozen 3B head) & \textbf{0.585} & \textbf{0.688} & \textbf{0.542} & \textbf{0.347} \\
Calpha-Joint (LLM + 100M head, co-trained) & 0.581 & 0.632 & 0.488 & 0.229 \\
Calpha-Token (DPLM-2 tokens) & 0.560 & 0.606 & 0.508 & 0.275 \\
ESM2 $\rightarrow$ SimpleFold-3B (fine-tuned) & 0.610 & 0.696 & 0.702 & 0.489 \\
ESMFold & 0.590 & 0.778 & 0.816 & 0.636 \\
SimpleFold-3B & 0.590 & 0.711 & 0.760 & 0.547 \\
SimpleFold-100M & 0.572 & 0.688 & 0.641 & 0.479 \\
OpenDDE (single sequence) & 0.414 & 0.392 & 0.308 & 0.162 \\
OpenDDE (+MSA)$^\dagger$ & 0.710 & 0.902 & 0.897 & 0.631 \\
ESM3-open (1.4B) & 0.599 & 0.664 & 0.619 & 0.527 \\
DPLM-2 (650M) & 0.561 & 0.545 & 0.435 & 0.303 \\
DPLM-2 Bit (650M) & 0.481 & 0.547 & 0.516 & 0.359 \\
MultiFlow$^\ddagger$ & 0.396 & 0.258 & 0.232 & 0.129 \\
BioMatrix-1.7B$^\ddagger$ & 0.230 & 0.175 & 0.138 & 0.073 \\
BioMatrix-4B$^\ddagger$ & 0.242 & 0.171 & 0.119 & 0.064 \\
Nearest training structure (copy) & 0.368 & 0.464 & 0.553 & 0.327 \\
Random walk (floor) & 0.162 & 0.137 & 0.137 & 0.110 \\
\bottomrule
\end{tabular}

\end{table}

\begin{table}[H]
\centering
\caption{TM-score by chain length, CAMEO-2025.}
\label{tab:len-cameo}
\footnotesize\renewcommand{\arraystretch}{0.93}
\begin{tabular}{@{}lcccc@{}}
\toprule
Model & $<$100 & 100--299 & 300--600 & $>$600 \\
 & \small(7) & \small(101) & \small(88) & \small(4) \\
\midrule
\textbf{Calpha} (LLM + frozen 3B head) & \textbf{0.811} & \textbf{0.929} & \textbf{0.897} & \textbf{0.387} \\
Calpha-Joint (LLM + 100M head, co-trained) & 0.783 & 0.915 & 0.793 & 0.324 \\
Calpha-Token (DPLM-2 tokens) & 0.735 & 0.841 & 0.777 & 0.363 \\
ESM2 $\rightarrow$ SimpleFold-3B (fine-tuned) & 0.828 & 0.929 & 0.921 & 0.666 \\
ESMFold & 0.839 & 0.911 & 0.912 & 0.737 \\
SimpleFold-3B & 0.831 & 0.939 & 0.954 & 0.719 \\
SimpleFold-100M & 0.795 & 0.913 & 0.898 & 0.555 \\
OpenDDE (single sequence) & 0.571 & 0.418 & 0.303 & 0.296 \\
OpenDDE (+MSA)$^\dagger$ & 0.894 & 0.963 & 0.960 & 0.894 \\
ESM3-open (1.4B) & 0.783 & 0.870 & 0.850 & 0.568 \\
DPLM-2 (650M) & 0.668 & 0.806 & 0.722 & 0.445 \\
DPLM-2 Bit (650M) & 0.758 & 0.848 & 0.823 & 0.551 \\
MultiFlow$^\ddagger$ & 0.381 & 0.303 & 0.230 & 0.188 \\
BioMatrix-1.7B$^\ddagger$ & 0.205 & 0.292 & 0.214 & 0.138 \\
BioMatrix-4B$^\ddagger$ & 0.205 & 0.363 & 0.270 & 0.099 \\
Nearest training structure (copy) & 0.762 & 0.835 & 0.868 & 0.659 \\
Random walk (floor) & 0.151 & 0.140 & 0.131 & 0.115 \\
\bottomrule
\end{tabular}

\end{table}

One pattern in Table~\ref{tab:validity} is worth stating in words.
Over 99\% of adjacent C$\alpha$ distances pass for both Calpha and the experimental structures on every set.
\calpha{}-Joint, with its small co-trained head, is regular up to 600 residues but not beyond:
above 600 residues only 4--29\% of its alpha-carbon distances are physical, a failure that its
TM-score alone would understate.

\section{Development Experiments}
\label{app:dev}

This appendix gives the experiments behind Section~\ref{sec:decomp}.
All development results are FoldBench evaluations with one prediction per scored target.
Figure~\ref{fig:ladder} labels each trained configuration with the date of its evaluation.
Random is a random-walk baseline with no learned parameters, and the randomly initialised
head (Scratch) is evaluated at step 1{,}667. The joint-training run in the figure (0.750)
predates the checkpoint reported as \calpha{}-Joint in Table~\ref{tab:main} (0.744).

\subsection{Replacing token output by a continuous head}
\label{sec:output}

At two billion training tokens, the early Qwen model generated ESM3 tokens by greedy decoding.
The mean across 333 scored chains was 0.5617. The \calpha{}-Token checkpoint used in the
main benchmark also saw two billion training tokens, but was trained separately to write
DPLM-2 tokens. It averaged 0.6623 on its 333 completed chains. Both token runs lacked
a score for \texttt{8uxt\_A}. Applying the main benchmark's zero-for-failure rule
gives 0.6603 across all 334 chains. The 0.562 and 0.660 labels therefore describe
different checkpoints, token representations and scoring populations.

For the 100M-head experiment we kept the ESM3-token backbone frozen and trained a pretrained SimpleFold-100M
head on its hidden states for 10{,}000 steps. Three controls received the same head
training recipe with different conditioning signals (Table~\ref{tab:output}).

\begin{table}[!htb]
\centering
\caption{Same head and recipe, different conditioning (FoldBench-334; SimpleFold-100M head,
10{,}000 steps). The first row is the same language model decoding through its own token
output (333 scored chains). The four flow-head rows cover 334 chains.}
\label{tab:output}
\small
\begin{tabular}{@{}lcc@{}}
\toprule
Conditioning of the head & TM & No hom. \\
\midrule
(none: the language model writes structure tokens) & 0.562 & -- \\
\midrule
Amino-acid one-hot (floor) & 0.346 & 0.375 \\
Base Qwen hidden states (no protein training) & 0.347 & 0.372 \\
Hidden states of the protein-trained Qwen & 0.712 & 0.677 \\
ESM2-3B features (control) & 0.789 & 0.723 \\
\bottomrule
\end{tabular}
\end{table}

On the 333 shared chains the ESM3-token and 100M-head means were 0.562 and 0.713. This
checks the output-path comparison on the same targets.
The head starts from the public SimpleFold-100M weights, so part of the gain may come from its
structural pretraining and not from the continuous output form.
With the 100M head, states from a Qwen without protein training yielded nearly the same
mean TM as one-hot sequence input (paired $\Delta$TM $+0.0003$, CI $[-0.007, +0.008]$).
In a separate 3B-head configuration, ESM2 features injected through a trained projector
gave 0.476 with a Qwen that had received no protein training. The input representation and head
size both matter when interpreting this comparison.

\subsection{Historical comparisons with ESM2-conditioned SimpleFold heads}
\label{sec:headsize}
On FoldBench-334, the 100M-head comparison gave a paired difference of $-0.077$ TM
(CI $[-0.090,-0.064]$). With the 3B head fine-tuned on both paths for 10,000 steps at
crop 256, the historical comparison gave $-0.031$ (CI $[-0.048,-0.015]$).
The final comparison in Table~\ref{tab:toll} includes the changes to training
listed in Table~\ref{tab:recipes}. Each interval pertains to its own configuration.

\subsection{Co-training and structural pretraining}

Training the language model jointly with the 100M head, instead of freezing it, raised TM from
0.712 to 0.750 (paired $+0.038$, CI $[+0.023, +0.053]$; $+0.052$ on the no-homolog bucket).
Two controls from the same experiment bound what the result depends on.
A head initialised at random instead of from the SimpleFold weights reached only 0.270 at the
step where the pretrained head reached 0.672. The pretrained head scored higher on all 334
targets at that checkpoint.
A run without ESM2 input was stopped early; its in-training probes were at 0.26--0.32 against
0.37--0.57 for the main run at matched steps, which is suggestive but not a formal result.
The stage-1 language evaluation is reported in Section~\ref{sec:onemodel}.

\subsection{Effect of unfreezing the pretrained head}
\label{sec:freeze}

\begin{figure}[t]
\centering
\includegraphics[width=0.82\linewidth]{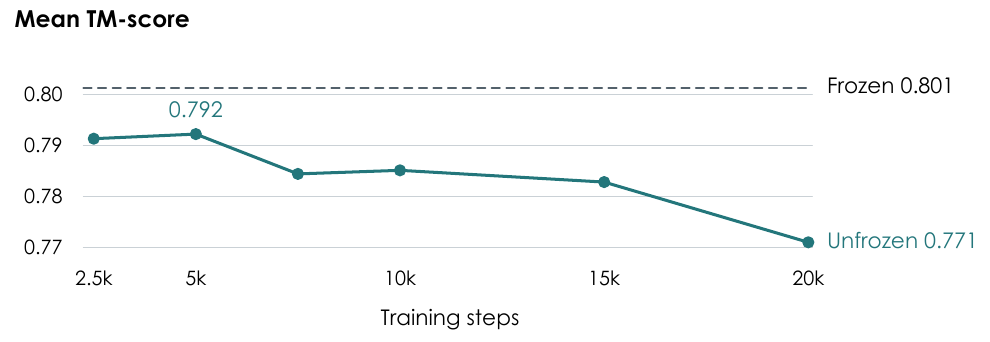}
\caption{Unfreezing the pretrained 3B head (FoldBench-334). The dashed line marks the starting frozen configuration. All six measured checkpoints score lower after unfreezing; the highest mean is 0.792 at 5,000 updates.}
\label{fig:unfreeze}
\end{figure}

Swapping in the 3B head and retraining it on our 84{,}658 chains gave 0.791.
Freezing it and training only the 8.4M-parameter front-end gave 0.801 (paired $+0.0105$,
CI $[+0.002, +0.020]$); this comparison also changed the stage-1 language model that supplies
the hidden states, so the gain cannot be attributed to freezing alone.
The clean test is the reverse experiment (Figure~\ref{fig:unfreeze}): starting from the frozen
configuration and unfreezing the head, accuracy was significantly below the frozen model at the
first checkpoint measured ($-0.010$ at 2{,}500 steps), peaked at 0.792, and fell to 0.771 by
20{,}000 steps.
The measured gap persists through the last checkpoint in this run.

\subsection{Crop length helps a little; several other things do not}
\label{sec:nothelp}

Table~\ref{tab:knobs} summarises the remaining experiments, each changing one setting and all starting from
the frozen-head configuration.

\begin{table}[!htb]
\centering
\caption{Experiments that change one setting of the frozen-head configuration (FoldBench-334, paired
$\Delta$TM). $^{*}$: 95\% CI excludes zero.}
\label{tab:knobs}
\small
\begin{tabular}{@{}llr@{}}
\toprule
Question & Change & $\Delta$TM \\
\midrule
Is the front-end under-trained? & 10k $\rightarrow$ 20k steps & $+0.0001$ \\
Is the front-end too small? & double its capacity & $-0.0027$ \\
Is the training crop too short? & 256 $\rightarrow$ 512 residues & $+0.0060^{*}$ \\
\quad same, with the language model unfrozen & 256 $\rightarrow$ 512 residues & $+0.0071^{*}$ \\
Should the head adapt? & unfreeze the 3B head, 10k steps & $-0.0162^{*}$ \\
Should the language model adapt? & unfreeze Qwen, crop 256 & $-0.0021$ \\
 & unfreeze Qwen, crop 512 & $-0.0011$ \\
\bottomrule
\end{tabular}
\end{table}

\paragraph{Crop length.}
Raising the training crop from 256 to 512 residues is the only change that helped, and the gain
falls where the diagnosis predicts: $+0.015$ on 300--600 residues and nothing below 300.
The effect was reproduced along two independent paths, with the language model frozen and
unfrozen.
It is small, about one fifth of the gap to the public head, and close to the size at which
retraining noise could matter; we have not measured that noise.

\paragraph{Unfreezing the language model.}
Letting gradients from the head update all 9B parameters of Qwen changed nothing overall and made
the longest bucket significantly worse ($-0.049$ above 600 residues at crop 512).
This experiment does not test whether additional long-chain training would help.

\paragraph{A recurring shape.}
Across these experiments, gains landed in the homolog bucket while the no-homolog bucket did not
move significantly (for the final configuration against the first frozen one: $+0.011^{*}$ with
homologs, $-0.002$ without).
These development experiments were all scored on FoldBench-334; as Section~\ref{sec:toll} shows,
the homolog pattern does not carry over to CAMEO-2025, so we report it as an observation on our
development set and not as an explanation.

\section{Reproduction Gates, Baseline Details and Reproducibility}
\label{app:repro}

\paragraph{Reproduction gates.}
Table~\ref{tab:gates} records, for each third-party model, the check we ran before using it and
its outcome.
For ESMFold and ESM3 the check is agreement between this pipeline and our earlier runs,
which establishes that the pipeline is repeatable but is not an external validation.

\begin{table}[H]
\centering
\caption{Third-party models: weights, training cut-off where documented, and the outcome
of the reproduction gate.}
\label{tab:gates}
\small
\begin{tabular}{@{}p{2.6cm}p{3.4cm}p{2.6cm}p{6.0cm}@{}}
\toprule
Model & Weights & Training cut-off & Gate \\
\midrule
ESMFold & \texttt{esmfold\_v1} & Unverified here & Repeatable: 0.8692 vs.\ archived 0.8697 (FoldBench-334) \\
ESM3-open & \texttt{esm3-sm-open-v1} & Unverified here & Repeatable: 0.7871 vs.\ archived 0.7880 \\
SimpleFold & public 100M / 3B & as released & Evaluated with our sampler; no external number checked \\
OpenDDE & \texttt{opendde\_v1} (PyPI 1.2.0) & 2021-09-30 & No external number checked. The official self-test passes; we ran torch 2.10.0, not the pinned 2.7.1, without an equivalence check \\
DPLM-2 (650M) & \texttt{dplm2\_650m} & not stated & Own CAMEO-2022 set: RMSD reproduced (7.59 vs.\ 7.7\,\AA); TM not (0.71 vs.\ 0.79) \\
DPLM-2 Bit (650M) & \texttt{dplm2\_bit\_650m} & not stated & RMSD reproduced (6.38 vs.\ 6.4\,\AA); TM not (0.77 vs.\ 0.84). The only weights released for the DPLM-2.1 paper; they implement its bit-wise token modelling only \\
MultiFlow & Zenodo 10714631 & not stated & Not run. Our batched sampler matches the official one on short chains (max alpha-carbon deviation 0.10\,\AA) \\
BioMatrix-1.7B/4B & \texttt{BioMatrix-*-SFT} & not stated & Not reproduced: 0.29 / 0.45 on its own test set vs.\ 0.84 reported \\
\bottomrule
\end{tabular}
\end{table}

\paragraph{BioMatrix.}
We prompt with the single instruction shared by all folding rows of its public fine-tuning data,
decode greedily, and convert the generated structure tokens to coordinates with the GCP-VQVAE
decoder.
We excluded an error on our side by decoding the structure tokens of its own training data with
the same decoder, which gives TM 0.993.
Two explanations remain open: 271 of the 449 sequences of the test set on which the paper
reports folding appear verbatim in the public fine-tuning data, and the paper may have evaluated
task-specific checkpoints while the public one is merged.
We have not contacted the authors.
Under our protocol only 24 of 334 FoldBench-334 generations have exactly the right number of
structure tokens.

\paragraph{OpenDDE.}
OpenDDE is run with its command-line defaults (five samples, top-ranked reported) with and
without an MSA.
The MSAs come from the public ColabFold server~\cite{mirdita2022colabfold} with no date filter on the sequence databases,
so the MSA row is an upper reference and not a held-out result.
The 155 CASP16 domains correspond to 72 distinct sequences; OpenDDE is run once per sequence.

\paragraph{Frontier chat models.}
The model identifiers are \texttt{gpt-6.1-sol}, \texttt{claude-opus-5-5},
\texttt{gemini-4-argon} and \texttt{deepseek-v4.1-flash}, queried on 2026-11-11 with default
reasoning settings and no tools.
The DeepSeek responses come from two endpoints serving the same model (87 targets through a
gateway, 93 through the vendor API).
Parseable output was returned for 180 (GPT), 172 (Claude), 179 (Gemini) and 176 (DeepSeek) of
the 180 targets; the rest score zero.

\paragraph{Eligible sequences for the trigger test.}
The trigger evaluation only forms prompts for sequences of at most 400 residues. This
leaves 278, 140 and 155 eligible entries for FoldBench-334, CASP16 and CAMEO-2025, and
these are the denominators of the rates in Table~\ref{tab:trigger}. The rates for the
fifth request type correspond to 273/278, 135/140 and 148/155.

\end{document}